\documentclass[a4paper]{article}
\usepackage[margin=25mm]{geometry}
\usepackage{graphicx}
\usepackage{subfigure}
\usepackage{float}
\usepackage{graphicx}
\usepackage{amssymb}
\usepackage{mathrsfs}
\usepackage{fullpage}
\usepackage{amsmath}
\usepackage{graphicx,caption}
\usepackage{eucal}
\usepackage{lipsum}
\usepackage{amsfonts}
\DeclareMathAlphabet{\mathpzc}{OT1}{pzc}{m}{it}
\usepackage{epsfig}
\usepackage{adjustbox}
\allowdisplaybreaks
\newtheorem{theorem}{Theorem}[section]
\usepackage{latexsym}
\usepackage[nodisplayskipstretch]{setspace}
\makeatletter
\def\ps@pprintTitle{%
\let\@oddhead\@empty
\let\@evenhead\@empty
\def\oddfoot\@empty
\let\evenfoot\@oddfoot}
\makeatother
\usepackage{verbatim}

\providecommand{\keywords}[1]
{
  \small	
  \textbf{\textit{Keywords---}} #1
}

\title{Cost analysis of a $M\!A\!P\!/P\!H/S$ performability model with $P\!H$ retrial times using simulated annealing method}
\author{ Vidyottama Jain$^{1}$, Raina Raj$^{1}$ and S. Dharmaraja$^{2}$  \\
        \small Central University of Rajasthan, Ajmer, India$^{1}$\\
       \small Indian Institute of Technology Delhi,  India$^{2}$\\
}
\date{} 

\begin{document}
\maketitle

\begin{abstract}
This work focuses over the   performability analysis  of  a multi-server   retrial queueing model  with phase-type inter-retrial times  in  cellular networks. It is considered that the  pattern of the new call arrival and handoff call arrival follows Markovian arrival process ($\textrm{\it MAP}$). The service times of both types of calls  are  phase-type ($\textrm{\it PH }$) distributed with different service rates, and inter-failure \& inter-repair times  of channels are exponentially distributed.   
For the prioritization of handoff calls,  $\mathpzc{G}$  channels are kept in reserve  for handoff calls.  When all the available channels, say $\mathpzc{S}$, are busy at the arrival epoch of a handoff call, the handoff call   will be dropped.  Whereas a new call will be blocked and will have an option to join the orbit of  infinite capacity or leave the system without getting the connection, if at least $\mathpzc{S}-\mathpzc{G}$ channels are busy.   A  new call in the orbit, termed as  retrial call,   retries to get the connection  after a random interval which follows  $\textrm{\it PH}$ distribution.  This model is  analyzed as a level-dependent-quasi-birth-death ($\textrm{\it LDQBD}$) process by applying matrix-analytic method ($\textrm{\it{MAM}}$). Further, the closed-form expressions for essential performance   measures of the proposed model are derived. Through  numerical illustrations, the  behaviour of  performance measures depending on the  various relevant intensities  is discussed. An expected cost optimization problem  is formulated to determine the optimal value of service intensity  and the optimal value of repair intensity. The cost function analysis is executed by employing  Simulated Annealing ($\textrm{\it{SA}}$) method.
\end{abstract} \hspace{10pt}

\keywords{Channel Failure,  LDQBD Process,  Markovian Arrival Process,  Performability Model, Phase-Type Distribution,  Retrial  Queue, Simulated Annealing.}

\section{Introduction} \label{section1}

The tremendous increment in the  number of mobile users and the
need for better network quality lead to a massive 
wireless cellular networks intensification. The consideration of user behaviour, in particular the repeated attempts (retrials) of users whose services have been denied due to the lack of available channels, is essential to determine the  performance of the systems. In wireless cellular networks, single-server and multi-server retrial queuing models, where   the incoming call arrival follows  the Poisson process and service times is exponentially distributed, have been extensively  studied (\cite{aguir2004,artalejo2010, van2011,jain2020}). In the cutting-edge wireless technologies, due to the burstiness  of the incoming calls, such type of arrival process will  not be able to capture correlated arrival times. To obtain  realistic performance parameters,  the arrival process of  input should be modelled  by the Markovian arrival process ($\textrm{\it MAP}$) and the service times distribution should be described by the phase-type ($\textrm{\it PH}$) distribution.

Over the last few decades, retrial models with  $\textrm{\it MAP}$ arrival and/or $\textrm{\it PH}$ service  distributions are investigated  by several researchers.  Some of the relevant studies with exponentially distributed inter-retrial times are discussed here.   \cite{diamond1998} dealt with $\textrm{\it MAP/PH/1}$  retrial queueing model.  \cite{artalejo2006} studied \textrm{\it MAP/M/$\mathpzc{S}$} retrial queueing model and determined the maximal  number of customers in orbit by applying matrix-analytic method ($\textrm{\it MAM}$).   \cite{dudina2013} proposed a $\textrm{\it MAP/PH/$\mathpzc{S}$}$ retrial queueing model with  finite buffer and analyzed its stationary distribution.   \cite{zhou2013} studied  a multi-server retrial queueing system where arrival times  and service times of  handoff and new calls are represented by different  $\textrm{\it MAP}$ and  different $\textrm{\it PH}$  distributions, respectively.  \cite{kim2014}  proposed a multi-server retrial queueing model considering the guard channel policy for the prioritization of handoff calls. They considered   exponentially distributed service times with different rates. Further,    \cite{dudin2016}  proposed a better version of (\cite{kim2014}) by assuming $\textrm{\it PH}$  distributed service times. 

In wireless networks, the inter-retrial times are notably brief in comparison to service times. Since, the retrial attempt is just a matter of pushing one button, these retrial customers will make numerous attempts during any given service interval. Therefore, the consideration of exponential retrial times in place of non-exponential ones could lead to under or over estimating the system parameters. Though the consideration of $\textrm{\it PH}$ distributed inter-retrial times can result in a complex model due to the exponential growth of the state space.  However,  for the steady-state analysis of the various systems, approximation and truncation methods have been applied in the literature.
A brief overview of such relevant works  is as follows. \cite{alfa2002} proposed  $\textrm{\it MAP/PH/$\mathpzc{S}$}$ queueing model with retrial phenomenon where inter-retrial times are described by $\textrm{\it PH}$ distribution.
Though the method, used in \cite{alfa2002} to  approximate the infinite capacity orbit, reduced  the computational complexities of the steady-state analysis, yet it failed to cope up with the real life scenario.    They also explored the behavior of a multi-server queueing model by considering general $\textrm{\it PH}$ distributions for service and inter-retrial times but completely disregarded the correlation behavior among the input flow by considering quasi-random input.
On the similar line, \cite{dharmaraja2008} presented explicit expression for the generator matrix of  $\textrm{\it MAP/PH/$\mathpzc{S}$}$ retrial queueing system for  voice, video, and data traffics.  \cite{shin2011} analyzed $\textrm{\it M/M/$\mathpzc{S}$}$  retrial queueing model using level-dependent-quasi-birth-death process $\textrm{\it (LDQBD)}$   by considering two-phase $\textrm{\it PH}$ distribution for retrials. Later on,  they proposed an   approximation for the distribution of the number of busy servers as well as the mean number of customers in retrial orbit for the same model with relaxation in two-state $\textrm{\it PH}$ distribution constraint for retrials.  Very recently,    \cite{chakravarthy2020}  discussed the similar type of   $\textrm{\it M/M/$\mathpzc{S}$}$  retrial queueing model in which  the rates of $\textrm{\it PH}$ distributed inter-retrial times depend on the threshold parameters and provided the comparison with the model proposed by \cite{shin} using simulation results.

  Performability models, which consider that the failure behavior of the system has an impact over its performance (\cite{ma2001,trivedi2003,jindal2006}), are the need of  modern wireless cellular networks. The system failure can be described as the server/channel failure at any time, and obviously,  it needs repair.  The concept of unreliable server/ channel failure in cellular networks is understood as follows. In cellular networks, radio spectrum consists of channels through which service is provided to customers. Due to various reasons, i.e., hardware error, software error, fading, noise, interference, etc., these channels may become unavailable.  This scenario is termed as channel failure.  When a channel failure occurs, the call, which is being served by that  channel is lost but  other calls will be smoothly carried out. Once the channel failure is detected, it will be considered for repair immediately (one such process is automatic software reconfiguration). 
  In the literature, some relevant queueing models with unreliable servers are as follows.
   \cite{yang2009} considered a $\textrm{\it M/PH/$\mathpzc{S}$}$  queueing system following Poisson process for failure of servers and  exponentially distributed repair times. 
  Recently,  \cite{dudin2019} proposed a multi-server retrial queueing model with $\textrm{\it MAP}$ arrival and phase-type with failure distribution ($\textrm{\it PHF}$)  for the service and failure times. They assumed exponentially distributed  inter-retrial times and developed a new algorithm to provide the  approximate  stationary distribution.

In this paper,  a  $\textrm{\it MAP/PH/$\mathpzc{S}$}$ retrial queueing model with orbit of  infinite capacity,  $\textrm{\it PH}$ distributed retrial times, prioritization of calls,  exponentially distributed channel failure times and   repair times is introduced. To the best of authors' knowledge, the proposed model is the first one that deals with such complex system. This is a homogeneous model where each cell is identically and independently distributed (i.i.d.) of each other. The incoming calls, categorized as handoff calls $(\mathcal{H})$ and new calls $(\mathcal{N})$, follow  $\textrm{\it MAP}$. The service time distribution of handoff call and new call are described by $\textrm{\it PH}$ distributions with different intensities. The guard channel policy is applied for the prioritization of  handoff calls.  As per this policy, $\mathpzc{G}$ $(0 \leq \mathpzc{G} < \mathpzc{S})$ channels are exclusively kept in reserve for handoff calls. Thus, the new calls, which find at least $\mathpzc{S}-\mathpzc{G}$ channels busy upon their arrival, will be blocked and will have an option  to join the orbit of  infinite capacity or leave the system without getting the connection.  New blocked calls after joining the orbit are referred as  retrial calls (refer, \cite{jain2020}).  The proposed unreliable system follows exponentially distributed failure times and repair times. With these assumptions, the stochastic behavior of this underlying model can be explained by the $\textrm{\it LDQBD}$ process. Since the analytic solution  of the stationary distribution for $\textrm{\it LDQBD}$ process is  intractable, the steady-state analysis is performed by using  $\textrm{\it MAM}$,  developed by   \cite{neuts1990}. Due to the large size of state space, the capacity of orbit is approximated by using a general method proposed by   \cite{bright1995} for the numerical computation purpose. More details over the  $\textrm{\it MAM}$ can be found in \cite{he2014}, \cite{latouche1999}  and \cite{neuts1994}. To minimize the cost per unit time for the service provider, a cost function has been developed and an efficient meta-heuristic method, named Simulated Annealing ($\textrm{\it SA}$) is employed to obtain the optimal service intensity and repair intensity to minimize the cost function.

This work is  arranged in six sections.  In Section \ref{section2},  a $\textrm{\it MAP/PH/$\mathpzc{S}$}$ model with  $\textrm{\it PH}$ distributed inter-retrial times and exponentially distributed  channel failures is mathematically demonstrated.  Section \ref{section3} describes the behaviour   of the  $\textrm{\it LDQBD}$ process. Also, steady-state probabilities are computed  through the $\textrm{\it MAM}$. In Section \ref{section4}, formulas of the key performance measures to analyse the network efficiency  are derived explicitly. Numerical  illustrations to point out the impact of various intensities over the system performance are presented in Section \ref{section5} along with the cost optimization problem. Finally, the underlying model is concluded with the insight for the future works in Section \ref{section7}.

\section{Model Description}\label{section2}
This work considers a multi-server retrial queuing model with channel failures and repairs in wireless cellular networks. Figure. \ref{fig:fig_sim1} demonstrates the outline of the  proposed model. All the important assumptions are provided as follows:
 
\begin{figure}
\centering
\includegraphics[width=5.9in, height=3.6in]{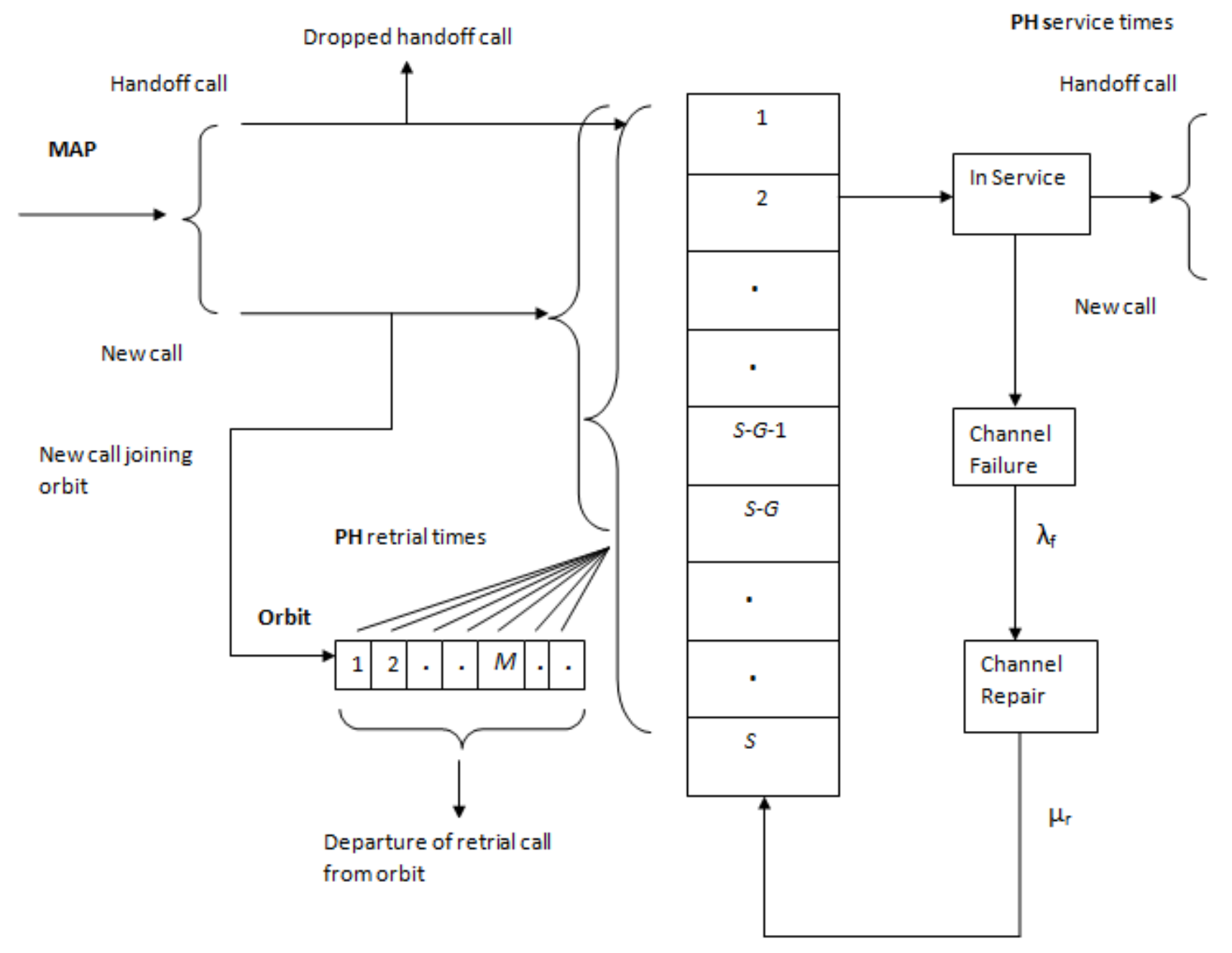} 
\caption{A multi-server performability  model with MAP, PH service and retrial times.}
\label{fig:fig_sim1}
\end{figure}

\begin{itemize}

\item Arrival Process:

\hspace{3mm} Arrival of handoff calls and new calls follows a continuous time   $\textrm{\it MAP}$  with dimension $L$. Let $C$ be the irreducible infinitesimal generator of this $\textrm{\it MAP}$ where $C = C_0 + C_{\mathcal{H}} + C_{\mathcal{N}}$. Let  $\pi$ be the unique solution of $\pi C = O$ and $\pi e =1$, where $O$ is a zero row vector of appropriate size and $e$ is a unit column vector. The  arrival intensities of the $\textrm{\it MAP}$ for  handoff call and new call are defined as  $\lambda_{\mathcal{H}} = \pi C_{\mathcal{H}}e$ and  $\lambda_{\mathcal{N}} = \pi C_{\mathcal{N}}e$, respectively.

\item {Service Process:} 

\hspace{3mm} The new call service times   in a cell    follows  the  $\textrm{\it PH}$ distribution  with the representation $(\delta_{\mathcal{N}}, L_{\mathcal{N}})$ and dimension $W_1$,  i.e., $L_{\mathcal{N}} e + L^{0}_{\mathcal{N}} =0$. Similarly,  the  handoff call service times  in a cell  following $\textrm{\it PH}$ distribution is  represented by $(\delta_{\mathcal{H}}, L_{\mathcal{H}})$ and dimension $W_1$,  i.e., $L_{\mathcal{H}} e + L^{0}_{\mathcal{H}} =0$. Note that notations $\oplus$ and  $\otimes$ are used for the Kronecker sum and  the   Kronecker product of two matrices, respectively. For more description over Kronecker sum and Kronecker product, authors suggest readers to  refer  \cite{dayar2012}.
\item Retrial Process:

\hspace{3mm} The retrial times of a retrial call  is $\textrm{\it PH}$ distributed with representation $(\gamma, \Gamma)$ and dimension $W_2$, i.e., $\Gamma e + \Gamma^0(1) + \Gamma^0(2)=0.$ Here $\Gamma^0(1)$ shows the absorption due to the departure from the cell and $\Gamma^0(2)$ denotes the absorption due to the retrial attempt. The fundamental retrial intensity is given by $1/ \theta = -\gamma (\Gamma)^{-1} e. $

\item Channel Failure and Repair:

\hspace{3mm} It has been proposed that the channel failure may occur while the system is in  busy state and will follow  exponential distribution with intensity $\lambda_f$. The failed channels will be  repaired immediately  and will follow exponential distribution with intensity $\mu_r$. 
\end{itemize}

\section{Mathematical Model}\label{section3}

The underlying stochastic process \{$\Xi(t), t \geq 0 \}$ for the cell is defined by the following state space:
\begin{align}
\nonumber \Omega &= \{(\mathpzc{l}, \kappa, \mathfrak{j},  \mathfrak{i}, v, s_{\mathcal{N}}, s_{\mathcal{H}}, \mathpzc{r}); \mathpzc{l} \geq 0,~0 \leq \kappa \leq \mathpzc{S},~ 0 \leq \mathfrak{i} \leq \mathpzc{S}, ~0 \leq \mathfrak{i}+\kappa \leq \mathpzc{S},0 \leq \mathfrak{j} \leq min\{\kappa, \mathpzc{S}-\mathpzc{G}\},1 \leq v \leq L\},\end{align}
 where, 
\begin{itemize}
\item $\mathpzc{l}$ denotes  the number of retrial calls,
\item $\kappa$ denotes the  number of busy channels,
\item $\mathfrak{j}$ denotes the number of ongoing new calls in the system, 
\item $\mathfrak{i}$ denotes the  number of failed channels,
\item $v$ is the current phase of $\textrm{\it MAP}$,
 \item $s_{\mathcal{N}}= \{(s_{\mathcal{N}}^1, s_{\mathcal{N}}^2,\ldots,s_{\mathcal{N}}^{\mathfrak{j}}); 1 \leq s_{\mathcal{N}}^{\nu_1} \leq W_1; 1 \leq \nu_1 \leq \mathfrak{j}\},$ where the $\nu_1^{\textrm{th}}$  new call,  out of  $\mathfrak{j}$ number of  new calls, is being served in phase $s_{\mathcal{N}}^{\nu_1},$
\item $s_{\mathcal{H}}= \{(s_{\mathcal{H}}^1, s_{\mathcal{H}}^2,\ldots,s_{\mathcal{H}}^{\kappa-\mathfrak{j}}); 1 \leq s_{\mathcal{H}}^{\nu_2} \leq W_1; 1 \leq \nu_2 \leq \kappa-\mathfrak{j}\},$ where the $\nu_2^{\textrm{th}}$  handoff call,  out of  $\kappa-\mathfrak{j}$ number of handoff calls, is being served in phase $s_{\mathcal{H}}^{\nu_2},$
\item $\mathpzc{r} = \{(r^1, r^2,\ldots, r^{\mathpzc{l}}); 1 \leq r^{h} \leq W_2; 1 \leq h \leq \mathpzc{l}\},$ where the $h^{\textrm{th}}$ retrial call  is in phase $r^h.$

\end{itemize}
The stochastic process \{$\Xi(t), t \geq 0 \}$ can be described as  $\textrm{\it LDQBD}$ process and its tri-diagonal infinitesimal generator matrix will be given as:\\
\begin{center}
$\mathscr{Q} =
\begin{pmatrix}
 \mathscr{Q}_{0,0} & \mathscr{Q}_{0,1} & 0 & 0 & 0 &  \\
     \mathscr{Q}_{1,0}& \mathscr{Q}_{1,1}& \mathscr{Q}_{1,2} & 0 & 0 &  \\
      0  & \mathscr{Q}_{2,1} & \mathscr{Q}_{2,2} & \mathscr{Q}_{2,3} & 0 &   \\
      &   &  & \ddots & \ddots & \ddots \\
     &   & & & \ddots & \ddots & \ddots 
      \end{pmatrix}.$\\
\end{center}

We define the following notations in order to carry out the further analysis. 
\begin{itemize}
\item $I_u$ denotes an identity matrix of order $u$.
\item O(A) denotes the order of matrix A.
\item $\Phi_{\mathcal{N}}(\mathfrak{j}) = \underbrace{L_{\mathcal{N}} \oplus L_{\mathcal{N}} \oplus \ldots \oplus L_{\mathcal{N}}}_{\mathfrak{j}}$ represents that $\mathfrak{j}$ number of new call are receiving service.

\item  $\Phi_{\mathcal{H}}(\kappa-\mathfrak{j}) = \underbrace{L_{\mathcal{H}} \oplus L_{\mathcal{H}} \oplus \ldots \oplus L_{\mathcal{H}}}_{\kappa-\mathfrak{j}}$ represents that $\kappa-\mathfrak{j}$ number of handoff calls are receiving service.

\item $ \Phi_{orbit}(\mathpzc{l}) = \underbrace{\Gamma \oplus \Gamma \oplus \ldots \oplus \Gamma}_{\mathpzc{l}}$ 
represents that $\mathpzc{l}$ number of  new calls are in the process of retrying.
\item $\hat \Phi_{orbit}(\mathpzc{l}) = \underbrace{(\Gamma^0(2)\gamma) \oplus (\Gamma^0(2)\gamma) \oplus \ldots \oplus (\Gamma^0(2)\gamma)}_{\mathpzc{l}}$ represents that $\mathpzc{l}$ number of new calls are having unsuccessful retrial attempt.  
\item $\Psi_{\mathcal{N}}(\mathfrak{j}) = \displaystyle{\sum_{y=0}^{\mathfrak{j}-1}I_{W_1^y} \otimes L_{\mathcal{N}}^0 \otimes I_{W_1^{\mathfrak{j}-y-1}}}$ represents  that any one out of the $\mathfrak{j}$ number of new calls has completed the service.

\item $\displaystyle{\Psi_{\mathcal{H}}(\kappa-\mathfrak{j}) = \sum_{y=0}^{\kappa-\mathfrak{j}-1}I_{W_1^y} \otimes L_{\mathcal{H}}^0 \otimes I_{W_1^{\kappa-\mathfrak{j}-y-1}}}$ represents that  any one out of the $\kappa-\mathfrak{j}$ number of handoff calls has completed the service.

\item $\Psi_{orbit}(\mathpzc{l}+1) = \displaystyle{\sum_{y=0}^{\mathpzc{l}-1}I_{W_2^y} \otimes \Gamma^0(1) \otimes I_{W_2^{\mathpzc{l}-y-1}}}$ represents that any one out of the $\mathpzc{l}$ number of retrial calls leaves the orbit as well as the cell without getting connected.

\item $\hat \Psi_{orbit}(\mathpzc{l}+1) = \displaystyle{\sum_{y=0}^{\mathpzc{l}-1}I_{W_2^y} \otimes (\Gamma^0(2) \otimes \delta_{\mathcal{N}}) \otimes I_{W_2^{ \mathpzc{l}-y-1}}}$ represents that any one out of the $\mathpzc{l}$ number of retrial calls is getting service after its  successful retrial. 
\item Following matrices have been used as the notation matrices to represent the block matrices of generator  $\mathscr{Q}$\\

\scalebox{0.8}{
$ F^{+} = \begin{pmatrix}
 0 & 1 & 0   & \cdots &0 &0 \\
  0& 0& 1 & \cdots  & 0 &0\\
 \vdots      & \vdots & \vdots & \ddots  & \vdots & \vdots\\
 0  & 0 & 0 & \cdots& 0&1\\
    0    &  0    &  0&     \hdots  &0 &0
     \end{pmatrix},$
$F^{-}=
\begin{pmatrix}
 1 &0  & \cdots  &0 & 0 \\
     0&  1 &  \cdots  &0&0  \\
   \vdots  & \vdots & \ddots  & \vdots &\vdots \\
  0 & 0 & \cdots & 0&1\\
0& 0& \hdots & 0&0
      \end{pmatrix}$,
  $H^{+} = \begin{pmatrix}
 0 & 1 & 0   & \cdots &0 &0 \\
  0& 0& 1 & \cdots  & 0 &0\\
 \vdots      & \vdots & \vdots & \ddots  & \vdots & \vdots\\
 0  & 0 & 0 & \cdots& 0&1\\
     \end{pmatrix},$ 
 $H^{-} = \begin{pmatrix}
 1 & 0 & 0   & \cdots &0 &0 \\
  0& 1& 0 & \cdots  & 0 &0\\
 \vdots      & \vdots & \vdots & \ddots  & \vdots & \vdots\\
 0  & 0 & 0 & \cdots& 1&0\\
     \end{pmatrix}.$
}
\end{itemize}
The intensities of the upper diagonal of the $\mathscr{Q}$ matrix represent the scenario when
 one new blocked call joins the orbit. The intensities of the lower diagonal show the loss of one retrial call either due to the successful retrial or due to the departure from orbit without obtaining the service. The main diagonal represents transitions due to the arrival or service of handoff calls and new calls or the transitions of retrial call from one phase to another phase. The number of retrial calls is  not  changed during these transitions.  The block matrices are defined as follows.
{ \small
 \begin{align*}
& \textbf{ Upper Diagonal :}\\
&\mathscr{Q}_{\mathpzc{l},\mathpzc{l}+1} = \text{diag}\{X_\mathpzc{l}(0), X_\mathpzc{l}(1), \ldots, X_\mathpzc{l}(S)  \};~~\mathpzc{l} \geq 0,\\
&X_\mathpzc{l}(\kappa)  = \text{diag}\{X_\mathpzc{l}(\kappa,0),X_\mathpzc{l}(\kappa,1), \ldots, X_\mathpzc{l}(\kappa,\mathpzc{b_\kappa}) \};~~ \kappa = \overline{0,\mathpzc{S}},~ \mathpzc{b_\kappa} =  min\{\kappa,\mathpzc{S}-\mathpzc{G}\},\\
 &X_\mathpzc{l}(\kappa,\mathfrak{j}) = \text{diag}\{X_\mathpzc{l}(\kappa,\mathfrak{j},0),X_\mathpzc{l}(\kappa,\mathfrak{j},0),\ldots,X_\mathpzc{l}(\kappa,\mathfrak{j},0)\}; O(X_\mathpzc{l}(\kappa,\mathfrak{j},0)) = (\mathpzc{S}-\kappa+1),\\
& X_\mathpzc{l}(\kappa,\mathfrak{j}, 0) =\begin{cases}
0; & \text{$~~\forall \kappa=\overline{0,\mathpzc{S}-\mathpzc{G}-1},~\mathfrak{j} = \overline{0,\mathpzc{b_\kappa}}$},\\
(C_{\mathcal{N}} \otimes I_{W_1^{\kappa}W_2^{\mathpzc{l}}}) \otimes \gamma; & \text{$~~\forall \kappa=\overline{\mathpzc{S}-\mathpzc{G},\mathpzc{S}},~\mathfrak{j} = \overline{0,\mathpzc{b_\kappa}}$.}   
  \end{cases}\\
& \text{{\bf Lower Diagonal :}}\\
& \mathscr{Q}_{\mathpzc{l}+1,\mathpzc{l}} =
\begin{pmatrix}
		Z_\mathpzc{l}(0) & \hat{Z}_\mathpzc{l}(0) & 0   & \cdots &0&0  \\
		0& Z_\mathpzc{l}(1)& \hat{Z}_\mathpzc{l}(1) & \cdots  & 0&0 \\
		\vdots      & \vdots & \vdots & \ddots  & \vdots& \vdots\\
		0  & 0 & 0 & \cdots& Z_\mathpzc{l}(\mathfrak{j}-1) &\hat{Z}_\mathpzc{l}(S-1)\\
		0    &  0    &  0&     \hdots  & 0 & Z_\mathpzc{l}(S)
	\end{pmatrix}; ~~\mathpzc{l} \geq 0,\\
& Z_\mathpzc{l}(\kappa)  = \text{diag}\{Z_\mathpzc{l}(\kappa,0),Z_\mathpzc{l}(\kappa,1),\ldots,Z_\mathpzc{l}(\kappa,\mathfrak{j})\};~~\forall \kappa = \overline{0,\mathpzc{S}},\mathfrak{j} = \overline{0,\mathpzc{b_\kappa}},\\ 
& Z_\mathpzc{l}(\kappa,\mathfrak{j})  = I_{\mathpzc{S}-\kappa+1} \otimes Z_\mathpzc{l}(\kappa,\mathfrak{j},1); ~Z_\mathpzc{l}(\kappa,\mathfrak{j},1)=I_{L W_1^{\kappa}} \otimes \Psi_{orbit}(\mathpzc{l}+1);~~ \forall \kappa = \overline{0,\mathpzc{S}},~\mathfrak{j} = \overline{0,\mathpzc{b_\kappa}}, \\ 
& \hat{Z}_\mathpzc{l}(\kappa) = \hat{Z}_\mathpzc{l}(\kappa,0) \otimes F^{+}, O(F^{+})  =\begin{cases}
    (\kappa+1)\times(\kappa+2); & \text{ $\forall \kappa  = \overline{0,\mathpzc{S}-\mathpzc{G}-1}$},\\
   (\mathpzc{S}-\mathpzc{G}+1); & \text{$\forall \kappa = \overline{\mathpzc{S}-\mathpzc{G},\mathpzc{S}-1}$},
  \end{cases}\\
& \hat{Z}_\mathpzc{l}(\kappa,\mathfrak{j}) =  \hat{Z}_\mathpzc{l}(\kappa,\mathfrak{j},1) \otimes F^{-}; \text{  O($F^{-}$) = $(\mathpzc{S}-\kappa+1)\times(\mathpzc{S}-\kappa),$}\\
 &  \hat{Z}_\mathpzc{l}(\kappa,\mathfrak{j},1) = I_{L W_1^{\kappa}} \otimes \hat{\Psi}_{orbit}(\mathpzc{l}+1);~~ \forall \kappa = \overline{0,\mathpzc{S}-1}~,\mathfrak{j} = \overline{0,\mathpzc{b_\kappa}}. \\
& \text{{\bf Main Diagonal :}}\\
	&\mathscr{Q}_{\mathpzc{l},\mathpzc{l}} =
	\begin{pmatrix}
		Y_\mathpzc{l}(0) & \hat{Y}_\mathpzc{l}(0) & 0   & \cdots &0&0  \\
		\overline{Y}_\mathpzc{l}(1)& Y_\mathpzc{l}(1)& \hat{Y}_\mathpzc{l}(1) & \cdots  & 0&0 \\
		\vdots      & \vdots & \vdots& \ddots  & \vdots& \vdots\\
		0  & 0 &  0 & \cdots& Y_\mathpzc{l}(S-1) &\hat{Y}_\mathpzc{l}(S-1)\\
		0    &  0 &   0&     \hdots  & \overline{Y}_\mathpzc{l}(S) & Y_\mathpzc{l}(S)
	\end{pmatrix}; ~~\mathpzc{l} \geq 0,\\
& Y_\mathpzc{l}(\kappa) = \text{diag}\{Y_\mathpzc{l}(\kappa,0),Y_\mathpzc{l}(\kappa,1), \ldots, Y_\mathpzc{l}(\kappa,\mathfrak{j})\};~~ \kappa = \overline{0,\mathpzc{S}}, \mathfrak{j}= \overline{0,\mathpzc{b_\kappa}},\\ 
& Y_\mathpzc{l}(\kappa,\mathfrak{j})=
\begin{pmatrix}
  Y_\mathpzc{l}(\kappa,\mathfrak{j},0)  & 0 & \cdots & 0  &0 \\
     \mu_r I & Y_\mathpzc{l}(\kappa,\mathfrak{j},1)  & \cdots & 0&0 \\
    \vdots    & \vdots &  \ddots &   \vdots & \vdots \\
    0   & 0 &  \cdots &   Y_\mathpzc{l}(\kappa,\mathfrak{j},\mathpzc{S}-\kappa-1)& 0 \\
      0   & 0&\cdots   &     \mu_r (\mathpzc{S}-\kappa) I     & Y_\mathpzc{l}(\kappa,\mathfrak{j},\mathpzc{S}-\kappa)
   \end{pmatrix},\\
 & Y_\mathpzc{l}(\kappa,\mathfrak{j},\mathfrak{i}) =\begin{cases}
C_0 \oplus  \Phi_{\mathcal{N}}(\mathfrak{j}) \oplus \Phi_{\mathcal{H}}(\kappa-\mathfrak{j}) \oplus \Phi_{orbit}(\mathpzc{l}) - \mathfrak{i} \mu_r I - \kappa\lambda_f I;  \text{$\forall  ~\kappa= \overline{0,\mathpzc{S}-\mathpzc{G}-1},\mathfrak{j} = 0,\mathpzc{b_\kappa},~0 \leq \mathfrak{i} < \mathpzc{S}-\kappa$},\\
C \oplus  \Phi_{\mathcal{N}}(\mathfrak{j}) \oplus \Phi_{\mathcal{H}}(\kappa-\mathfrak{j}) \oplus \Phi_{orbit}(\mathpzc{l})- \mathfrak{i} \mu_r I - \kappa\lambda_f I;   \text{$\forall \kappa= \overline{0,\mathpzc{S}-\mathpzc{G}-1},\mathfrak{j} = 0,\mathpzc{b_\kappa}, ~\mathfrak{i}=\mathpzc{S}-\kappa$,}\end{cases}\\
 & Y_\mathpzc{l}(\kappa,\mathfrak{j},\mathfrak{i}) =\begin{cases}
C_0 \oplus  \Phi_{\mathcal{N}}(\mathfrak{j}) \oplus \Phi_{\mathcal{H}}(\kappa-\mathfrak{j}) \oplus \Phi_{orbit}(\mathpzc{l}) - \mathfrak{i} \mu_r I - 2\kappa\lambda_f I;  \text{$\forall \kappa= \overline{0,\mathpzc{S}-\mathpzc{G}-1}, 0 < \mathfrak{j} < \mathpzc{b_\kappa},~0 \leq \mathfrak{i} < \mathpzc{S}-\kappa,$}\\
C \oplus  \Phi_{\mathcal{N}}(\mathfrak{j}) \oplus \Phi_{\mathcal{H}}(\kappa-\mathfrak{j}) \oplus \Phi_{orbit}(\mathpzc{l})- \mathfrak{i} \mu_r I - 2\kappa\lambda_f I;   \text{$\forall \kappa= \overline{0,\mathpzc{S}-\mathpzc{G}-1}, 0 < \mathfrak{j} < \mathpzc{b_\kappa}, ~\mathfrak{i}=\mathpzc{S}-\kappa$,}\\
\end{cases}\\
 & Y_\mathpzc{l}(\kappa,\mathfrak{j},\mathfrak{i}) =\begin{cases}
C_0 \oplus  \Phi_{\mathcal{N}}(\mathfrak{j}) \oplus \Phi_{\mathcal{H}}(\kappa-\mathfrak{j}) \oplus \Phi_{orbit}(\mathpzc{l}) - \mathfrak{i} \mu_r I - \kappa\lambda_f I;  \text{$  \kappa = \mathpzc{S}-\mathpzc{G},\mathfrak{j} = 0,\mathpzc{b_\kappa},~ 0 \leq \mathfrak{i} < \mathpzc{S}-\kappa$},\\
(C_0+C_{\mathcal{H}}) \oplus  \Phi_{\mathcal{N}}(\mathfrak{j}) \oplus \Phi_{\mathcal{H}}(\kappa-\mathfrak{j}) \oplus \Phi_{orbit}(\mathpzc{l})- \mathfrak{i} \mu_r I - \kappa\lambda_f I;  \text{$  \kappa= \mathpzc{S}-\mathpzc{G},\mathfrak{j} = 0,\mathpzc{b_\kappa}, ~\mathfrak{i}=\mathpzc{S}-\kappa$,}\\
\end{cases}\\
 & Y_\mathpzc{l}(\kappa,\mathfrak{j},\mathfrak{i}) =\begin{cases}
C_0 \oplus  \Phi_{\mathcal{N}}(\mathfrak{j}) \oplus \Phi_{\mathcal{H}}(\kappa-\mathfrak{j}) \oplus \Phi_{orbit}(\mathpzc{l}) - \mathfrak{i} \mu_r I - 2\kappa\lambda_f I;  \text{$ \kappa = \mathpzc{S}-\mathpzc{G},0 < \mathfrak{j} < \mathpzc{b_\kappa},~0 \leq \mathfrak{i} < \mathpzc{S}-\kappa$},\\
(C_0+C_{\mathcal{H}}) \oplus  \Phi_{\mathcal{N}}(\mathfrak{j}) \oplus \Phi_{\mathcal{H}}(\kappa-\mathfrak{j}) \oplus \Phi_{orbit}(\mathpzc{l})- \mathfrak{i} \mu_r I - 2\kappa\lambda_f I;   \text{$ \kappa=\mathpzc{S}-\mathpzc{G}, 0 < \mathfrak{j} < \mathpzc{b_\kappa}, ~\mathfrak{i}=\mathpzc{S}-\kappa$,}\\
\end{cases}\\
  & Y_\mathpzc{l}(\kappa,\mathfrak{j},\mathfrak{i}) =\begin{cases}
C_0 \oplus  \Phi_{\mathcal{N}}(\mathfrak{j}) \oplus \Phi_{\mathcal{H}}(\kappa-\mathfrak{j}) \oplus \Phi_{orbit}(\mathpzc{l}) - \mathfrak{i} \mu_r I - \kappa\lambda_f I;  \text{$\forall \kappa= \overline{\mathpzc{S}-\mathpzc{G}+1,\mathpzc{S}-1}, \mathfrak{j} = 0,~ 0 \leq \mathfrak{i} < \mathpzc{S}-\kappa$},\\
(C_0+C_{\mathcal{H}}) \oplus  \Phi_{\mathcal{N}}(\mathfrak{j}) \oplus \Phi_{\mathcal{H}}(\kappa-\mathfrak{j}) \oplus \Phi_{orbit}(\mathpzc{l})- \mathfrak{i} \mu_r I - \kappa\lambda_f I;   \text{$\forall \kappa= \overline{\mathpzc{S}-\mathpzc{G}+1,\mathpzc{S}-1},\mathfrak{j} =0, ~\mathfrak{i}=\mathpzc{S}-\kappa$,}\\
\end{cases}\\
 & Y_\mathpzc{l}(\kappa,\mathfrak{j},\mathfrak{i}) =\begin{cases}
C_0 \oplus  \Phi_{\mathcal{N}}(\mathfrak{j}) \oplus \Phi_{\mathcal{H}}(\kappa-\mathfrak{j}) \oplus \Phi_{orbit}(\mathpzc{l}) - \mathfrak{i} \mu_r I - 2\kappa\lambda_f I;  \text{$\forall \kappa= \overline{\mathpzc{S}-\mathpzc{G}+1,\mathpzc{S}-1},\mathfrak{j}= \overline{1,\mathpzc{b_\kappa}},~ 0 \leq \mathfrak{i} < \mathpzc{S}-\kappa$},\\
(C_0+C_{\mathcal{H}}) \oplus  \Phi_{\mathcal{N}}(\mathfrak{j}) \oplus \Phi_{\mathcal{H}}(\kappa-\mathfrak{j}) \oplus \Phi_{orbit}(\mathpzc{l})- \mathfrak{i} \mu_r I - 2\kappa\lambda_f I; 
\text{$\forall \kappa= \overline{\mathpzc{S}-\mathpzc{G}+1,\mathpzc{S}-1},\mathfrak{j}= \overline{1,\mathpzc{b_\kappa}}, \mathfrak{i}=\mathpzc{S}-\kappa$,}\\ 
\end{cases}\\
 & Y_\mathpzc{l}(\mathpzc{S},0,0)  = (C_0 + C_{\mathcal{H}}) \oplus \Phi_{\mathcal{N}}(\mathfrak{j}) \oplus \Phi_{\mathcal{H}}(\kappa-\mathfrak{j}) \oplus \Phi_{orbit}(\mathpzc{l})  + I_{L W_1^{\mathpzc{S}}}
 \otimes \hat{\Phi}_{orbit}(\mathpzc{l}) - \mathfrak{i} \mu_r I- \mathpzc{S}\lambda_f I,\\
 & Y_\mathpzc{l}(\mathpzc{S},\mathfrak{j},0) = (C_0 + C_{\mathcal{H}}) \oplus \Phi_{\mathcal{N}}(\mathfrak{j}) \oplus \Phi_{\mathcal{H}}(\kappa-\mathfrak{j}) \oplus \Phi_{orbit}(\mathpzc{l})  + I_{L W_1^{\mathpzc{S}}}
 \otimes \hat{\Phi}_{orbit}(\mathpzc{l})
 - \mathfrak{i} \mu_r I- \mathpzc{S}j\lambda_f I;~~\forall \mathfrak{j}= \overline{1,\mathpzc{b_\kappa}}. \\
 & \hat{Y}_\mathpzc{l}(\kappa) =
\begin{pmatrix}
  \hat{Y}_\mathpzc{l}(\kappa,0,\mathcal{H}) & \hat{Y}_\mathpzc{l}(\kappa,0,\mathcal{N}) & \cdots& 0  &0  \\
    0 & \hat{Y}_\mathpzc{l}(\kappa,1,\mathcal{H}) & \cdots& 0  & 0 \\
   \vdots   & \vdots  & \ddots& \vdots & \vdots \\
    0   & 0  & \cdots& \hat{Y}_\mathpzc{l}(\kappa,\kappa,\mathcal{N}) & 0 \\
        0 & 0  & \cdots & \hat{Y}_\mathpzc{l}(\kappa,\kappa+1,\mathcal{H})&  \hat{Y}_\mathpzc{l}(\kappa,\kappa+1,\mathcal{N})
       \end{pmatrix},\\
       & O(\hat{Y}_\mathpzc{l}(\kappa))=\begin{cases}
    (\kappa+1)\times(\kappa+2); \text{ $\forall \kappa = \overline{0,\mathpzc{S}-\mathpzc{G}-1}$},\\
  (\mathpzc{S}-\mathpzc{G}+1);  \text{$\forall \kappa = \overline{\mathpzc{S}-\mathpzc{G},\mathpzc{S}-1}$},
  \end{cases}\\
&  \hat{Y}_\mathpzc{l}(\kappa,\mathfrak{j},\mathcal{H}) = \hat{Y}_\mathpzc{l}(\kappa,\mathfrak{j},\mathcal{H},\mathcal{H}) \otimes F^{-}; ~~ \hat{Y}_\mathpzc{l}(\kappa,\mathfrak{j},\mathcal{N}) = \hat{Y}_\mathpzc{l}(\kappa,\mathfrak{j},\mathcal{N},\mathcal{N}) \otimes F^{-}; O(F^{-}) = (\mathpzc{S}-\kappa+1)\times(\mathpzc{S}-\kappa),\\
&  \hat{Y}_\mathpzc{l}(\kappa,\mathfrak{j},\mathcal{H},\mathcal{H})  = ((C_{\mathcal{H}} \otimes I_{W_1^{\kappa}})\otimes \delta_{\mathcal{H}}) \otimes I_{W_2^{\mathpzc{l}}} ;~~ \forall \kappa = \overline{0,\mathpzc{S}-1},~\mathfrak{j} = \overline{0,\mathpzc{b_\kappa}},\\
& \hat{Y}_\mathpzc{l}(\kappa,\mathfrak{j},\mathcal{N},\mathcal{N}) =\begin{cases}
((C_{\mathcal{N}} \otimes I_{W_1^{\mathfrak{j}}})  \otimes \delta_{\mathcal{N}}) \otimes I_{W_1^{\kappa-\mathfrak{j}}}\otimes I_{W_2^\mathpzc{l}}; 
& \text{$\forall \kappa = \overline{0,\mathpzc{S}-\mathpzc{G}-1},~\mathfrak{j} = \overline{0,\mathpzc{b_\kappa}}$},\\
0; 
& \text{$\forall \kappa = \overline{\mathpzc{S}-\mathpzc{G},\mathpzc{S}-1},~\mathfrak{j} = \overline{0,\mathpzc{b_\kappa}}$.}   
  \end{cases} \\
  & \overline{Y}_\mathpzc{l}(\kappa)=
\begin{pmatrix}
 \overline{Y}_\mathpzc{l}(\kappa,0,\mathcal{H}) &0  & \cdots  & 0 &0 \\
      \overline{Y}_\mathpzc{l}(\kappa,1,\mathcal{N})& \overline{Y}_\mathpzc{l}(\kappa,1,\mathcal{H})    &\cdots & 0&0 \\
   \vdots & \vdots &   \ddots  & \vdots & \vdots\\
     0 & 0  &  \cdots  & \overline{Y}_\mathpzc{l}(\kappa,\kappa-1,\mathcal{H})&0\\
     0 & 0  & \cdots & \overline{Y}_\mathpzc{l}(\kappa,\kappa,\mathcal{N})&  \overline{Y}_\mathpzc{l}(\kappa,\kappa,\mathcal{H})
       \end{pmatrix},\\
       & O(\overline{Y}_\mathpzc{l}(\kappa))=\begin{cases}
    (\kappa+1)\times \kappa;  \text{ $\forall \kappa = \overline{1,\mathpzc{S}-\mathpzc{G}}$},\\
     (\mathpzc{S}-\mathpzc{G}+1);  \text{$\forall \kappa = \overline{\mathpzc{S}-\mathpzc{G}+1,\mathpzc{S}}$},
  \end{cases}\\
& \overline{Y}_{\mathpzc{l}}(\kappa,\mathfrak{j},\mathcal{H}) = \overline{Y}_{\mathpzc{l}}(\kappa,\mathfrak{j},\mathcal{H},\mathcal{H}) \otimes H^{-} + \kappa\lambda_f I \otimes H^{+}; ~~\overline{Y}_{\mathpzc{l}}(\kappa,\mathfrak{j},\mathcal{N}) = \overline{Y}_{\mathpzc{l}}(\kappa,\mathfrak{j},\mathcal{N},\mathcal{N}) \otimes H^{-} + \kappa\lambda_f I \otimes H^{+},\\
 &  O(H^{-}) = ~O(H^{+}) = (\mathpzc{S}-\kappa+1)\times (\mathpzc{S}-\kappa+2),\\ 
& \overline{Y}_\mathpzc{l}(\kappa,\mathfrak{j},\mathcal{H},\mathcal{H})  = I_{L W_1^{\mathfrak{j}}} \otimes  \Psi_{\mathcal{H}}(\kappa-\mathfrak{j}) \otimes I_{W_2^{\mathpzc{l}}}; \forall \kappa = \overline{1,\mathpzc{S}},~\mathfrak{j} = \overline{0,\mathpzc{b_\kappa}},\\
&\overline{Y}_\mathpzc{l}(\kappa,\mathfrak{j},\mathcal{N},\mathcal{N})  = I_{L W_1^{\kappa-\mathfrak{j}}} \otimes  \Psi_{\mathcal{N}}(\mathfrak{j}) \otimes I_{W_2^{\mathpzc{l}}}; \forall \kappa = \overline{1,\mathpzc{S}},~\mathfrak{j} = \overline{1,\mathpzc{b_\kappa}}.
\end{align*}
    }%
It can be observed that  a closed-form analytical solution  of the steady-state  distribution is intractable for the $\textrm{\it LDQBD}$ process. We now describe the algorithmic procedure adopted here.

\subsection{Steady-State Analysis} \label{subsection1}
 
Let ${z_s}$, partitioned as ${z_s} = \{{z_s}(0), {z_s}(1),\ldots, {z_s}(\mathpzc{M}-1), {z_s}(\mathpzc{M}),\ldots\}$, be the steady-state probability vector of  $\mathscr{Q}$, i.e.,
 \begin{align}
{z_s} \mathscr{Q} = 0; {z_s} e =1. \label{eq:a1}
\end{align}

 The components of  ${z_s}$ are denoted as ${z_s}(\mathpzc{l}) = \{{z_s}(\mathpzc{l},0),~ {z_s}(\mathpzc{l},1), \ldots, {z_s}(\mathpzc{l},\kappa)\},$  ${z_s}(\mathpzc{l},\kappa) = \{{z_s}(\mathpzc{l}, \kappa,0)$ $,\ldots, {z_s}(\mathpzc{l},\kappa,\mathfrak{j})\}$ and ${z_s}(\mathpzc{l},\kappa,\mathfrak{j}) = \{{z_s}(\mathpzc{l}, \kappa, \mathfrak{j},0),\ldots, {z_s}(\mathpzc{l},\kappa,\mathfrak{j}, \mathfrak{i})\};$ $~ \mathpzc{l}\geq 0,~ 0\leq \kappa \leq \mathpzc{S},~\mathfrak{j} = min\{ \kappa, \mathpzc{S}-\mathpzc{G}\},~ 0 \leq  \mathfrak{i} \leq \mathpzc{S}-\kappa.$ Here, 
 $\textrm{\it MAM}$ has been applied to solve the system of equations,  mentioned in  \eqref{eq:a1}. According to the algorithm provided by   \cite{neuts1994},  $z_s$ satisfies the matrix-geometric relationship ${z_s}(\mathpzc{l}+1) = {z_s}(\mathpzc{l}) \Re^{(\mathpzc{l})};~~\mathpzc{l} \geq 0$, where the family of the matrices $\{\Re^{(\mathpzc{l})},~ \mathpzc{l} \geq 0\}$, called rate matrices,  are the minimal non-negative solutions to the following system of equations
\begin{align}
\mathscr{Q}_{\mathpzc{l}-1,\mathpzc{l}} + \Re^{(\mathpzc{l})} \mathscr{Q}_{\mathpzc{l},\mathpzc{l}} + \Re^{(\mathpzc{l})}(\Re^{(\mathpzc{l}+1)} \mathscr{Q}_{\mathpzc{l}+1,\mathpzc{l}}) &= 0;~ \mathpzc{l} \geq 0.\label{eq:a2}
\end{align}

From the structure of the $\mathscr{Q}$ matrix, it is clear that the order of the block matrices  will create computational complexity while evaluating system performance measures. Thus, it is required to approximate the infinite capacity of the orbit. In the literature, various methods have been developed to approximate the size of orbit. We have used a general approach for $\textrm{\it LDQBD}$ system given  by  \cite{bright1995} which provides a cut-off value, say $\mathpzc{M}$, for the orbit capacity. 
Once a truncation level $\mathpzc{M}$ is determined, the original system $\mathscr{Q}$ is approximated by the new system with orbit of size $\mathpzc{M}$. Thus, the unique stationary distribution for the system exist and
the algorithm for computing $z_s$ works as follows:\\

\textbf{Algorithm:} 
\begin{itemize}
	\item Choose $\mathpzc{M}$ a large finite number such that $\displaystyle{\sum_{\kappa=0}^{\mathpzc{S}}\sum_{\mathfrak{j}=0}^{\mathpzc{S}-\mathpzc{G}}\sum_{ \mathfrak{i}=0}^{\mathpzc{S}-\kappa}{z_s}(\mathpzc{M},\kappa,\mathfrak{j}, \mathfrak{i})e < \epsilon}$, where $\epsilon>0$ is a pre-defined tolerance value.
	\item For $\mathpzc{l}= \mathpzc{M}, \mathpzc{M}-1,\ldots,1,$ compute and store\\
	$ \Re^{(\mathpzc{l}-1)}  = - \mathscr{Q}_{\mathpzc{l}-1,\mathpzc{l}}(\mathscr{Q}_{\mathpzc{l},\mathpzc{l}}+ \Re^{(\mathpzc{l})} \mathscr{Q}_{\mathpzc{l}+1,\mathpzc{l}})^{-1}.$
	\item Determine value of ${x_s}(0)(\mathscr{Q}_{0,0} + \Re^{(0)} \mathscr{Q}_{1,0})=0.$ 
	\item For $\mathpzc{l}= \mathpzc{M}-1, \mathpzc{M}-2,\ldots,1,0,$ compute ${x_s}(\mathpzc{l}+1) = {x_s}(\mathpzc{l}) \Re^{(\mathpzc{l})}.$
	\item By renormalizing $x_s = (x_s(0),x_s(1),\ldots,x_s(\mathpzc{M})),$ determine $z_s$.
\end{itemize}
The following  theorem shows the positive recurrence behavior of the $\textrm{\it LDQBD}$ process. Interested readers can refer \cite{latouche1999} for the proof of the Theorem \ref{theorem1}.

 \begin{theorem}\label{theorem1}
The $\textrm{\it LDQBD}$  is positive recurrent if and only if there exist a strictly positive solution to the system 
\begin{align}
\nonumber {z_s}(0)(\mathscr{Q}_{0,0} + \Re^{(0)}\mathscr{Q}_{1,0}) =0,
\end{align}
normalized by
\begin{align}
\nonumber {z_s}(0) \Big( \sum_{\mathpzc{l}=0}^{\mathpzc{M}} \prod_{\kappa=0}^{\mathpzc{l}-1} \Re^{(\kappa)} \Big )e =1.
\end{align}
The steady-state probability vector is given by
\begin{align}
\nonumber {z_s}(\mathpzc{l}) = {z_s}(0) \Big(  \prod_{\kappa=0}^{\mathpzc{l}-1} \Re^{(\kappa)} \Big )e;~ 0 \leq \mathpzc{l} \leq \mathpzc{M}.
\end{align} 
\end{theorem}

\section{Performance Measures}\label{section4}
 The following relevant  performance measures for  the proposed system  may be calculated, after computing the  steady-state distribution $z_s$.
\begin{enumerate}
\item The probability that there are $\mathfrak{j}$ number of new calls receiving service:
\[P_{\mathcal{N}}(\mathfrak{j}) = \sum_{\mathpzc{l}=0}^{\mathpzc{M}-1} \sum_{\kappa=1}^{\mathpzc{S}}{z_s} (\mathpzc{l},\kappa,\mathfrak{j},0)e + \sum_{\mathpzc{l}=0}^{\mathpzc{M}-1} \sum_{\kappa=1}^{\mathpzc{S}-1} \sum_{\mathfrak{i}=1}^{\mathpzc{S}-\kappa} {z_s} (\mathpzc{l},\kappa,\mathfrak{j}, \mathfrak{i})e. \]
\item The probability that there are $\mathfrak{j}^{'}$ number of handoff calls
receiving service:
\[P_{\mathcal{H}}(\mathfrak{j}^{'}) = \sum_{\mathpzc{l}=0}^{\mathpzc{M}-1}\sum_{\kappa=1}^{\mathpzc{S}}\sum_{\mathfrak{i}=0}^{\mathpzc{S}-\kappa} {z_s} (\mathpzc{l},\kappa,\kappa-\mathfrak{j}^{'}, \mathfrak{i})e.\]
\item Expected number of busy channels in the system:
   \[ E\!B = \sum_{\mathfrak{j}=0}^{\mathpzc{S}-\mathpzc{G}}~\mathfrak{j} P_{\mathcal{N}}(\mathfrak{j}) + \sum_{\mathfrak{j}^{'}=1}^{\mathpzc{S}}~\mathfrak{j}^{'}P_{\mathcal{H}}(\mathfrak{j}^{'}).\]
   \item The probability that there are $\mathpzc{l}$ number of retrial calls:
\[P_{orbit}(\mathpzc{l}) = \sum_{\kappa=0}^{\mathpzc{S}}\sum_{\mathfrak{j}=0}^{\mathpzc{S}-\mathpzc{G}}\sum_{ \mathfrak{i}=0}^{\mathpzc{S}-\kappa} {z_s} (\mathpzc{l},\kappa,\mathfrak{j}, \mathfrak{i})e. \]
\item Expected number of retrial calls:
 \[ E\!R = \sum_{\mathpzc{l}=0}^{\mathpzc{M}-1}\mathpzc{l}~P_{orbit}(\mathpzc{l}).\]  
\item Expected number of failed channels in the system:
   \[ E\!N
   = \sum_{\mathpzc{l}=0}^{\mathpzc{M}-1}\sum_{\kappa=0}^{\mathpzc{S}-1}\sum_{\mathfrak{j}=0}^{\mathpzc{S}-\mathpzc{G}}\sum_{ \mathfrak{i}=1}^{\mathpzc{S}-\kappa}  \mathfrak{i} ~{z_s} (\mathpzc{l},\kappa,\mathfrak{j}, \mathfrak{i})e.\] 
 \item The dropping probability of a handoff call:
\[ P_d =  \frac{1}{\lambda_{\mathcal{H}}} \Big(\sum_{\mathpzc{l}=0}^{\mathpzc{M}-1}\sum_{\mathfrak{j}=0}^{\mathpzc{S}-\mathpzc{G}}{z_s} (\mathpzc{l},\mathpzc{S},\mathfrak{j},0)\times C_{\mathcal{H}} e + \sum_{\mathpzc{l}=0}^{\mathpzc{M}-1} \sum_{\kappa=0}^{\mathpzc{S}-1}\sum_{\mathfrak{j}=0}^{\mathpzc{S}-\mathpzc{G}}{z_s} (\mathpzc{l},\kappa,\mathfrak{j},\mathpzc{S}-\kappa) \times C_{\mathcal{H}} e\Big). \]
\item The blocking probability of a new call:
\[P_b = \frac{1}{\lambda_{\mathcal{N}}} \Big(\sum_{\kappa=\mathpzc{S}-\mathpzc{G}}^{\mathpzc{S}}\sum_{\mathfrak{j}=0}^{\mathpzc{S}-\mathpzc{G}}{z_s} (\mathpzc{M}-1,\kappa,\mathfrak{j},0)\times C_{\mathcal{N}} e + \sum_{\mathpzc{l}=0}^{\mathpzc{M}-1} \sum_{\kappa=0}^{\mathpzc{S}-1}\sum_{\mathfrak{j}=0}^{\mathpzc{S}-\mathpzc{G}}{z_s} (\mathpzc{l},\kappa,\mathfrak{j},\mathpzc{S}-\kappa) \times C_{\mathcal{N}} e\Big).\]
\item The probability that a new call will complete the cell residence time without getting the service:

\[P_{leave}^{no-service} = \frac{1}{\theta}  \Big(\sum_{\mathpzc{l}=1}^{\mathpzc{M}-1}\sum_{\kappa=0}^{\mathpzc{S}}\sum_{\mathfrak{j}=0}^{\mathpzc{S}-\mathpzc{G}} \sum_{ \mathfrak{i}=0}^{\mathpzc{S}-\kappa}\mathpzc{l}~{z_s} (\mathpzc{l},\kappa,\mathfrak{j}, \mathfrak{i})  \otimes \Gamma^0(1) e\Big). \]
\item The intensity of output flow of successfully served handoff calls:  

\[ \lambda_{\mathcal{H}}^{out} =   \sum_{\mathpzc{l}=0}^{\mathpzc{M}-1}\sum_{\kappa=1;\mathfrak{j}\neq \kappa}^{\mathpzc{S}}\sum_{\mathfrak{j}=0}^{\mathpzc{S}-\mathpzc{G}} \sum_{ \mathfrak{i}=0}^{\mathpzc{S}-\kappa}{z_s} (\mathpzc{l},\kappa,\mathfrak{j}, \mathfrak{i})  \otimes L^{0}_{\mathcal{H}} e. \]
\item The probability of arbitrary type call loss due to the occurrence of channel failure: 
 \[ P_{loss}^{c-failure}( \mathfrak{i}) = \sum_{\mathpzc{l}=0}^{\mathpzc{M}-1}\sum_{\kappa=0}^{\mathpzc{S}- \mathfrak{i}}\sum_{\mathfrak{j}=0}^{\mathpzc{S}-\mathpzc{G}} {z_s} (\mathpzc{l},\kappa,\mathfrak{j}, \mathfrak{i})e.\] 
 \item Intensity by which a retrial call is successfully connected to an available channel:
 \[ \theta_r^{succ} =  \sum_{\mathpzc{l}=0}^{\mathpzc{M}-1}\sum_{\kappa=1}^{\mathpzc{S}}\sum_{\mathfrak{j}=0}^{\mathpzc{S}-\mathpzc{G}}\sum_{ \mathfrak{i}=0}^{\mathpzc{S}-\kappa} \theta {z_s} (\mathpzc{l},\kappa,\mathfrak{j}, \mathfrak{i})\otimes \Gamma^{0}(2)\otimes \delta_{\mathcal{N}}e.\]
  \item  The probability of channels availability in busy state: 
  \[ P_{c-avail} = \sum_{\mathpzc{l}=0}^{\mathpzc{M}-1}\sum_{\kappa=1}^{\mathpzc{S}}\sum_{\mathfrak{j}=0}^{\mathpzc{S}-\mathpzc{G}}\sum_{ \mathfrak{i}=0}^{\mathpzc{S}-\kappa}{z_s} (\mathpzc{l},\kappa,\mathfrak{j}, \mathfrak{i})e.\]
\item The total expected carried traffic:
\[ E\!C = \sum_{\mathpzc{l}=0}^{\mathpzc{M}-1}\sum_{\kappa=0}^{\mathpzc{S}}\sum_{\mathfrak{j}=0}^{\mathpzc{S}-\mathpzc{G}}\sum_{ \mathfrak{i}=0}^{\mathpzc{S}-\kappa} (\mathpzc{l}+\kappa){z_s} (i,\kappa,\mathfrak{j}, \mathfrak{i})e.\]
\end{enumerate}

The next task is to illustrate the behaviour of the key performance measures and to explore the impact of various intensities over the proposed model.

\section{Numerical Illustration}\label{section5}
The main motivation of this section is to analysis the qualitative behaviour of the proposed model. In this section, various scenarios of the proposed model are numerically analyzed by considering  different parameters of arrival, service and retrial processes, i.e., $\textrm{\it MAP}$, $\textrm{\it PH}$ distributions, Poisson, exponential, etc. For the service provider,  among all the other factors, service intensity and repair intensity are crucial factors as they determine the cost criterion for any type of unreliable system. It's not the optimal choice to make service intensity and repair intensity either too low or too high as it might lead to under / over utilization of system resources. Thus, optimal values of these parameters need to be obtained in order to minimize the expected cost for the system.  This leads to the formulation of a cost optimization problem concerning expected cost per unit time for the proposed model on the basis of the system performance measures mentioned in the Section \ref{section4}.

For the numerical illustration, we have assumed that there are $\mathpzc{S}=5$ identical channels and $\mathpzc{G}=3$ guard channels in the specific cell throughout the section. Due to the assumption of the $\textrm{\it PH}$ distribution of inter-retrial times, computational problems inevitably arise for larger values of $\mathpzc{M}$. Thus  the  truncation approach has been applied to compute $\mathpzc{M}$ by considering the  tolerance value $\epsilon = 10^{-5}$. The computational complexity  has been dealt by considering sparse block matrices. The failure intensity and repair intensity are set to $\lambda_f = 0.5$ and $\mu_r =1$, respectively. Let matrices for the $\textrm{\it MAP}$ be defined as follows
\begin{align}
 \nonumber
  C_{0}= \begin{pmatrix}
 -1.3431 & 0.0230\\
 0 & -17.183
\end{pmatrix},~~~C_{\mathcal{H}}= \begin{pmatrix}
 0.6600 & 0\\
 0.2567 & 8.3351
\end{pmatrix},~~~ C_{\mathcal{N}}= \begin{pmatrix}
 0.6600 & 0\\
 0.2567 & 8.3351
\end{pmatrix}.
\end{align}

 The  correlation coefficient of  arrival times is $C_{r}=0.2211$ and the  variation coefficient of arrival times is $C_{v}= 12.33.$ The average arrival intensity of a handoff call $(\lambda_{\mathcal{H}})$ and a new call $(\lambda_{\mathcal{N}})$ is 1. The total arrival intensity ($\lambda$) is provided by $ \lambda_{\mathcal{H}} +\lambda_{\mathcal{N}}$.
 
Let $\textrm{\it PH}$ distributions' parameters for the service intensities of a handoff and a new call  be
\begin{align}
\nonumber
\delta_{\mathcal{H}}= \begin{pmatrix}
 0.9, &0.1
\end{pmatrix},~~ L_{\mathcal{H}}= \begin{pmatrix}
 -1.999 & 1.99\\
 0 & -0.999
\end{pmatrix} \text{and}~~\delta_{\mathcal{N}}= \begin{pmatrix}
 0,& 1
\end{pmatrix}, ~~ L_{\mathcal{N}}= \begin{pmatrix}
 -1 & 1\\
 0 & -1
\end{pmatrix}, \text{respectively}.
\end{align}

The fundamental service intensity of handoff calls $(\mu_{\mathcal{H}})$  and the fundamental service intensity of new calls $(\mu_{\mathcal{N}})$ are 1.85 and 1, respectively. The total service intensity ($\mu$) is given by $ \mu_{\mathcal{H}} +\mu_{\mathcal{N}}$. The inter-retrial intensity of a retrial call, following  $\textrm{\it PH}$ distribution, is given by the parameters
\begin{align}
 \nonumber
 \gamma= \begin{pmatrix}
 0.5, & 0.5
\end{pmatrix},~~~ \Gamma = \begin{pmatrix}
 -2 & 2\\
 0 & -2
\end{pmatrix}.
 \end{align}
The average retrial intensity $(\theta)$ is  1.33. 
Additionally,  various queueing models are defined as  particular cases of the proposed  model in the Table \ref{tab:table1}.
\begin{table}[H]
	\centering
\begin{tabular}{|l|l|l|l|l|l|}
	\hline
	Model Name & Arrival Process & Service Process & Retrial Process & Failure process & Repair Process\\
		\hline
		Case I & $\textrm{\it MAP}$ & $\textrm{\it PH}$ & $\textrm{\it PH}$ & Exponential & Exponential\\
		\hline
		Case II & $\textrm{\it MAP}$ & $\textrm{\it PH}$ & Exponential & Exponential & Exponential\\
		\hline
		Case III & $\textrm{\it MAP}$ & Exponential & Exponential & Exponential & Exponential\\
		\hline
		Case IV & Poisson & Exponential & $\textrm{\it PH}$ & Exponential & Exponential\\
		\hline
		Case V & Poisson & Exponential & Exponential & Exponential & Exponential\\
		\hline
	\end{tabular}
	\caption{Various cases of the proposed queueing model.}
	\label{tab:table1}
\end{table}

To demonstrate the feasibility of the developed model,
we numerically describe some interesting observations of the proposed system through four numerical illustrations. These illustrations will present the behaviour of some performance measures with respect to the several intensities i.e., arrival, service, retrial, repair and failure intensities.\\

\noindent \textbf{Illustration 1:} The impact of service intensity of handoff call ($\mu_{\mathcal{H}}$) and service intensity of new calls ($\mu_{\mathcal{N}}$), over the expected number of busy channels in the system ($E\!B$),  the expected number of retrial calls  ($E\!R$) and flow intensity  of handoff calls that receive service successfully ($\lambda^{out}_{\mathcal{H}}$), is in focus.

\begin{figure*}
\centering
\subfigure[ $E\!B$  versus  $\mu_{\mathcal{N}}$]
{\includegraphics[trim= 2.5cm 0.1cm 2cm 0.5cm, clip=true, height = 5.3cm,width = 0.5\textwidth]{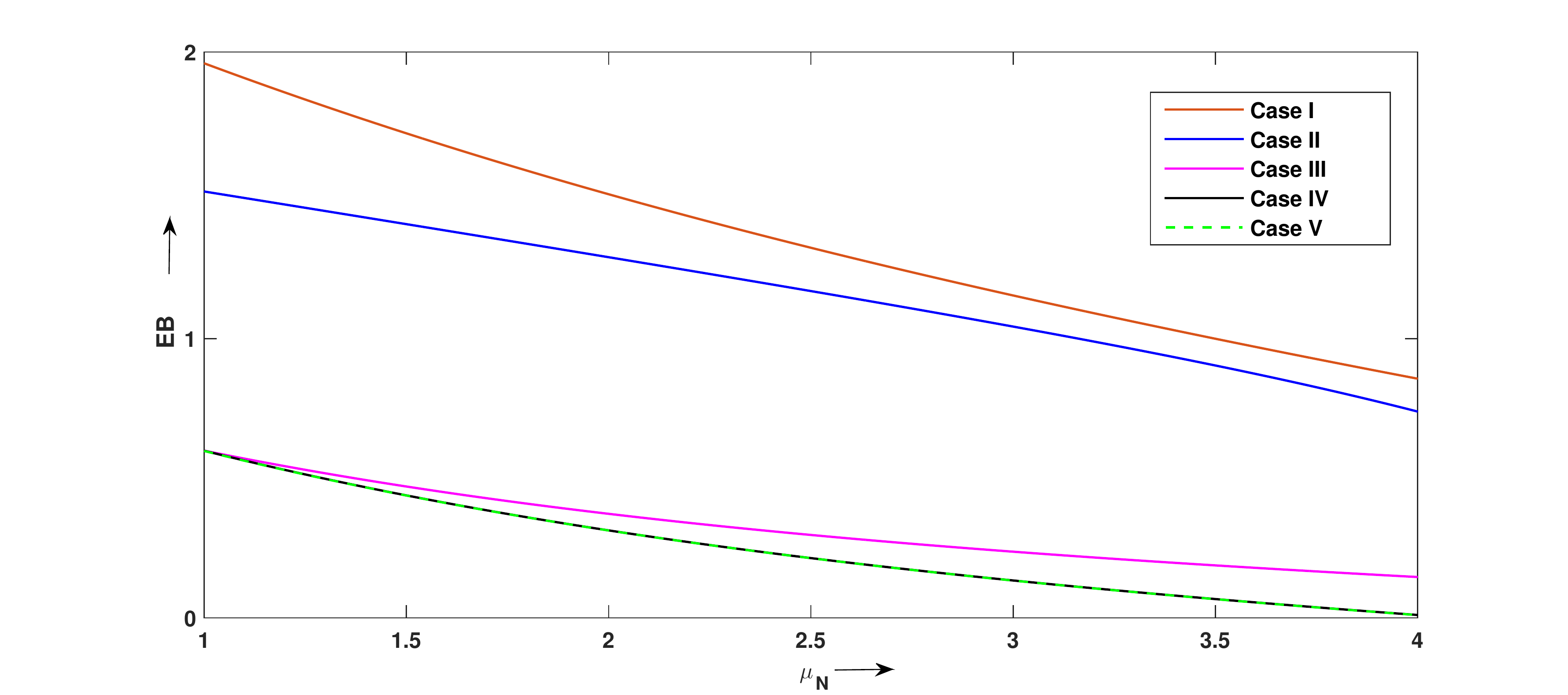}
\label{fig:EB}}%
\subfigure[ $E\!R$  versus  $\mu_{\mathcal{N}}$]
{\includegraphics[trim= 2cm 0.1cm 3cm 0.2cm, clip=true, height = 5.2cm,width = 0.5\textwidth]{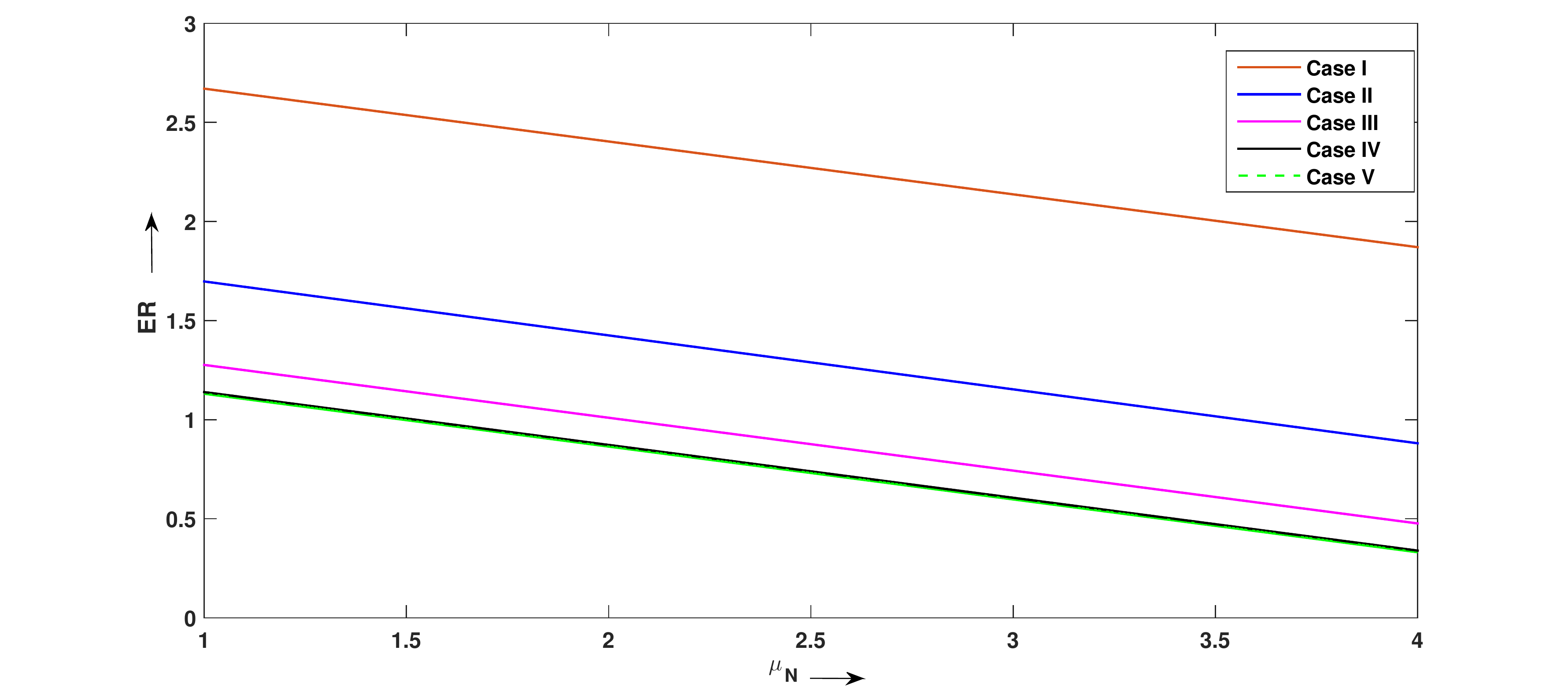}
\label{fig:ER}}%
\caption{Dependence of the expected number of busy channels in the system $E\!B$ and the expected number of retrial calls in the orbit $E\!R$
 over service intensity of a new call $\mu_{\mathcal{N}}$.}
\label{fig:EB_ER}
\end{figure*}

\begin{figure*}
\centering
{\includegraphics[ height =5.8cm,width = 0.6\textwidth]{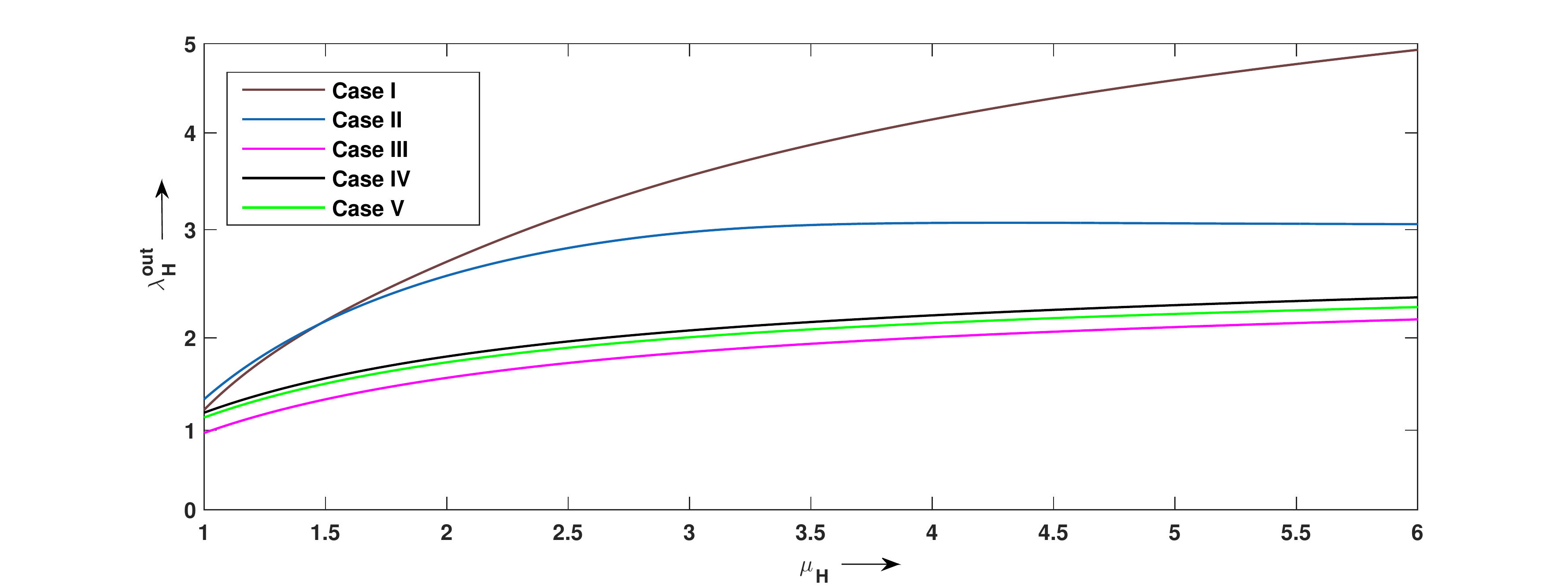}
}%
\caption{Dependence of the intensity of output flow of successfully served handoff calls $\lambda^{out}_{\mathcal{H}}$ over service intensity of a handoff call $\mu_{\mathcal{H}}$.}
\label{fig:rate_output_flow}
\end{figure*}

\begin{figure*}
\centering
\subfigure[$P_d$ versus  $\lambda_{\mathcal{H}}$]
{\includegraphics[trim= 3cm 0.1cm 3.8cm 0.4cm, height = 5.3cm,width = 0.5\textwidth]{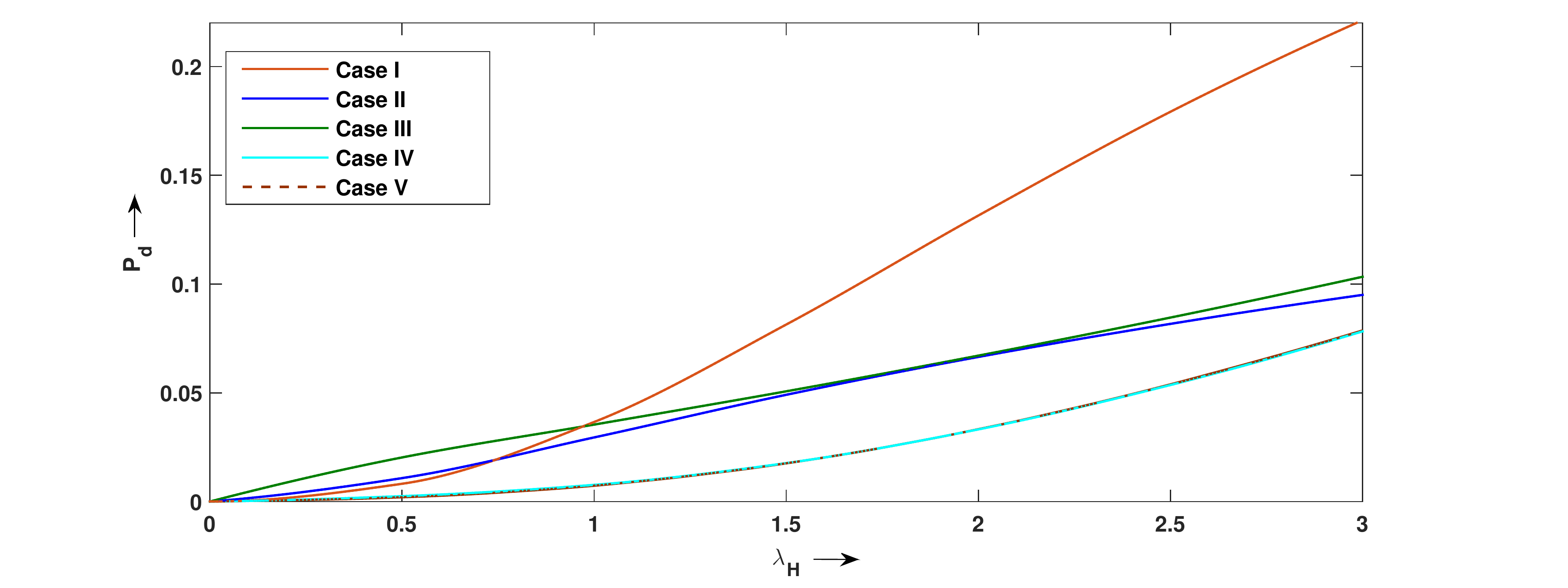}
\label{fig:Pd}}%
\subfigure[$P_b$ versus  $\lambda_{\mathcal{H}}$]
{\includegraphics[trim= 2.5cm 0.05cm 2cm 0.3cm, clip=true, height = 5.3cm,width = 0.5\textwidth]{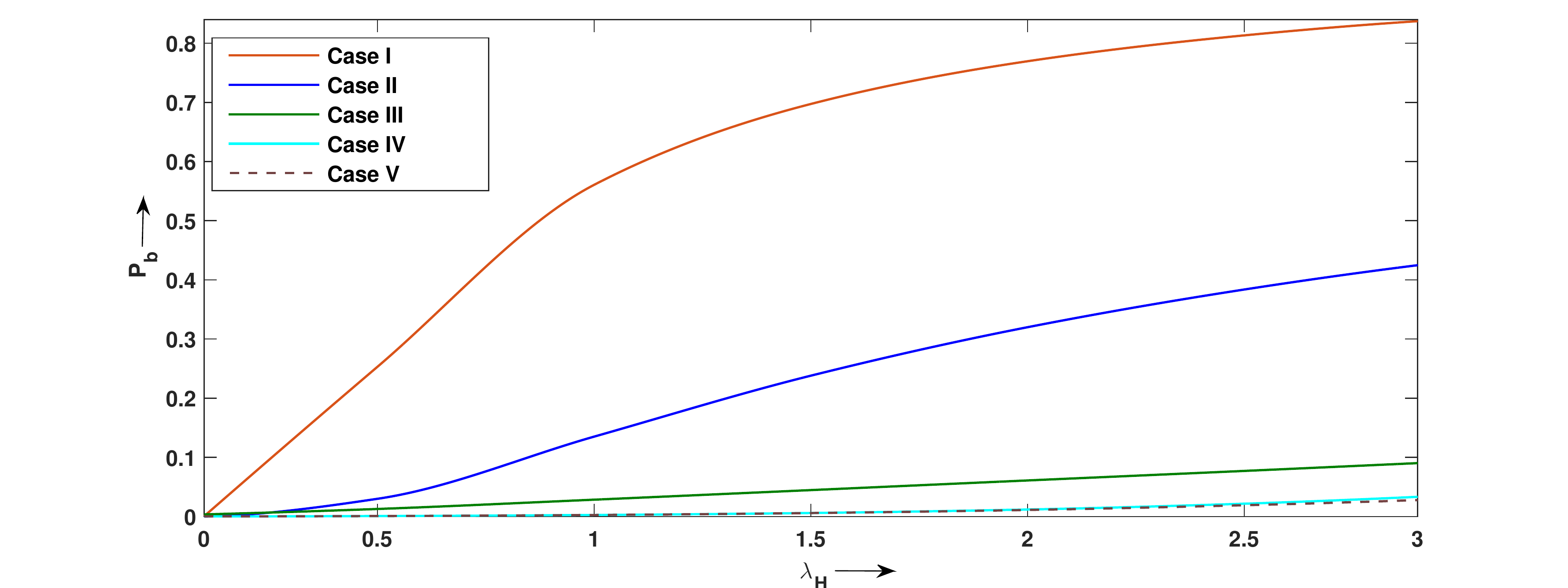}
\label{fig:Pb}}%
\caption{Dependence of the dropping probability $P_d$ and blocking probability $P_b$ over arrival intensity of a handoff call $\lambda_{\mathcal{H}}$. }
\label{fig:Pb_Pd}
\end{figure*}

\begin{itemize}
    \item Figures.  \ref{fig:EB_ER} \subref{fig:EB} and \ref{fig:EB_ER} \subref{fig:ER} show the dependencies of  $E\!B$ and  $E\!R$  over $\mu_{\mathcal{N}}$. It can be observed from the graphs that $E\!B$ and  $E\!R$  decrease with respect to $\mu_{\mathcal{N}}.$  An intuitive explanation for this finding could be given as follows. If $\mu_{\mathcal{N}}$ is increasing rapidly, calls are served at a faster speed which will reduce the number of retrial calls and busy channels in the system. One can also conclude from the graphs that the values of $E\!B$ and  $E\!R$  are significantly high for Case I in contrast to other cases which show that the correlation and the variation have a profound impact on $E\!B$ and  $E\!R$. 
   
    \item  Figure. \ref{fig:rate_output_flow} demonstrates that  $\lambda^{out}_{\mathcal{H}}$ increases as the handoff calls are served with increasing intensity $\mu_{\mathcal{H}}$. If the handoff calls are served with an increasing speed, it will result in increasing the number of handoff calls which are successfully served.  Moreover, the impact of  $\lambda^{out}_{\mathcal{H}}$ increases with the increasing value of variation and correlation coefficient for the service, retrial and arrival times, respectively.
   
\end{itemize}
\textbf{Illustration 2:} The objective here is to observe the effect of  arrival intensity of handoff calls ($\lambda_{\mathcal{H}}$) over the dropping probability ($P_d$) and blocking probability ($P_b$).
\begin{itemize}
    \item $P_d$  and $P_b$ have always been considered as the most essential performance measures in  cellular networks. Figures. \ref{fig:Pb_Pd} \subref{fig:Pd} and \ref{fig:Pb_Pd} \subref{fig:Pb} depict the behaviour of  $P_d$  and $P_b$ ($\lambda_{\mathcal{H}}$). $P_d$ and  $P_b$ appear to increase as a function of $\lambda_{\mathcal{H}}$. This is obvious as an increment in the value of $\lambda_{\mathcal{H}}$ will overload the system for the  fixed number of total channels. Thus, more calls will be lost from the system, resulting in a higher value of $P_d$ and  $P_b$. We also noticed that the value of $P_d$ and  $P_b$ are rapidly increasing  for the Case I as compare to other cases. This finding shows that  the consideration of Poisson arrival along with exponential service and retrial times can lead to major errors while evaluating the system performance in  cellular networks.
\end{itemize}
  
\begin{figure*}
\centering
\subfigure[$P_{loss}^{c-failure}$ versus  $\lambda_f$]
{\includegraphics[trim= 2cm 0.1cm 2.5cm 0.4cm, height = 5.3cm,width = 0.5\textwidth]{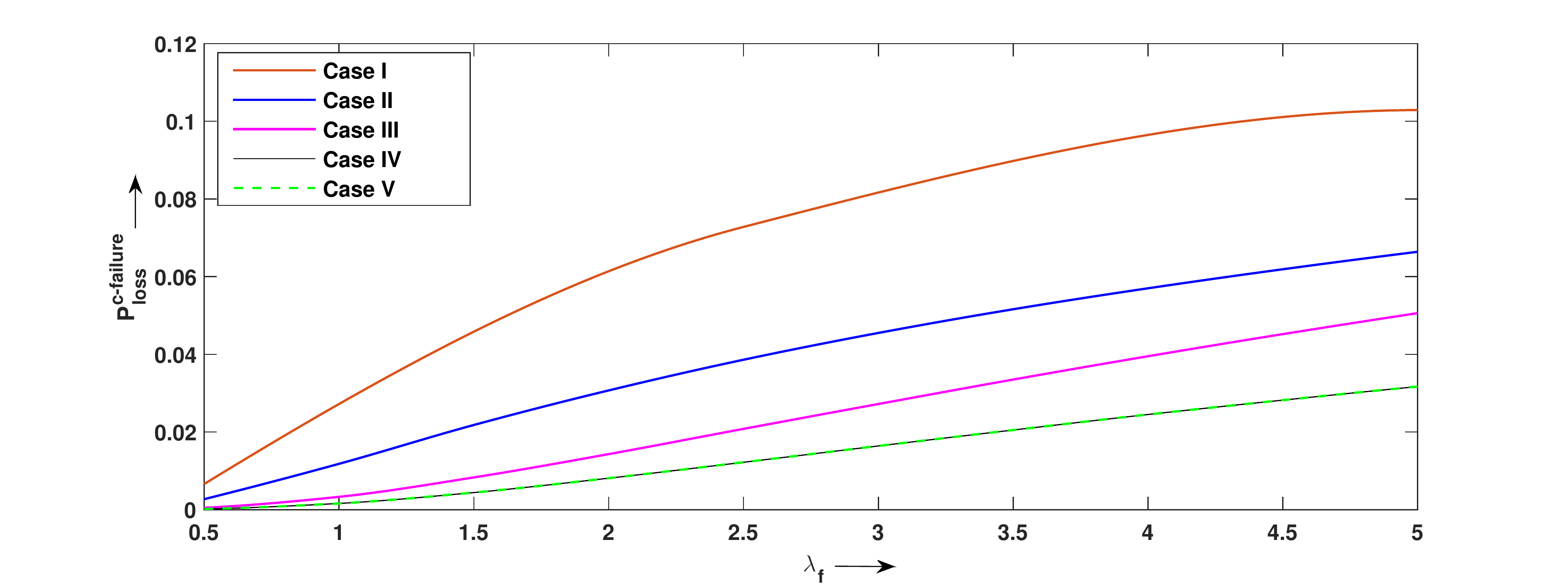}
\label{fig:loss_failure}}%
\subfigure[$P_{loss}^{c-failure}$ versus  $\mu_r$]
{\includegraphics[trim= 2cm 0.2cm 2cm 0.5cm, clip=true, height = 5.3cm,width = 0.5\textwidth]{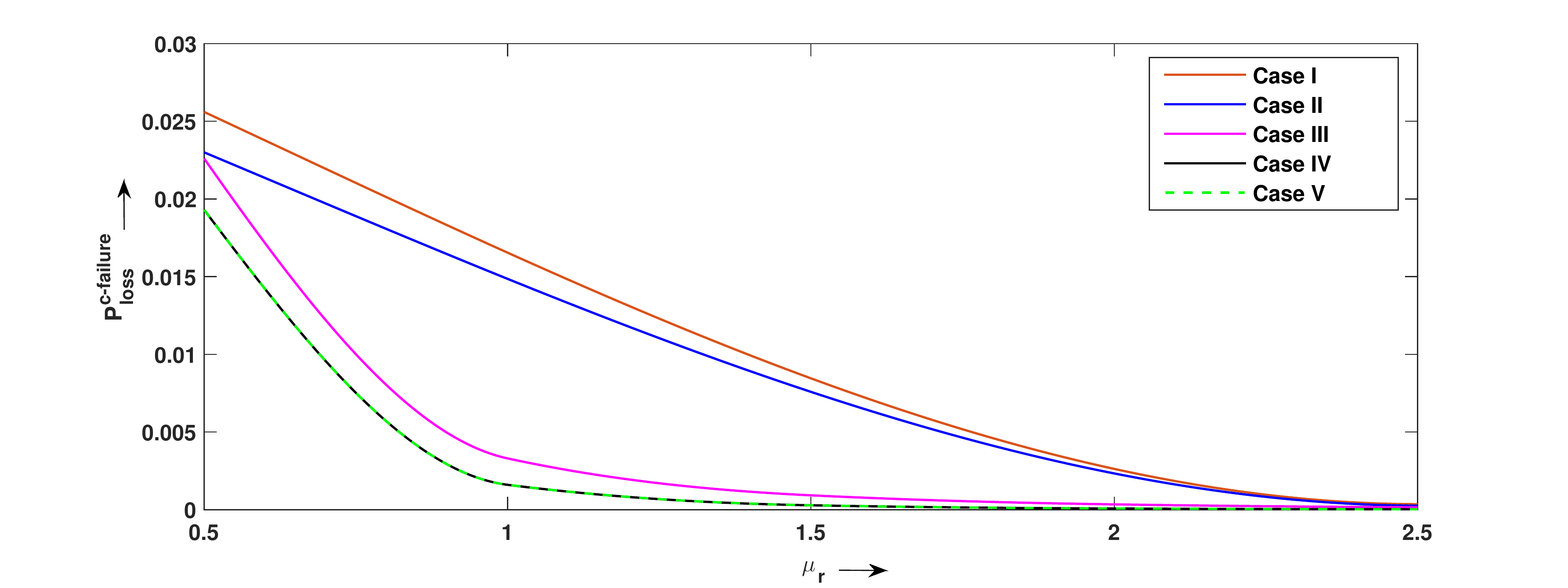}
\label{fig:loss_repair}}%
\caption{Dependence of the call loss probability  due to the occurrence of channel failure $P_{loss}^{c-failure}$ over failure intensity $\lambda_f$ and repair intensity $\mu_r$.}
\label{fig:P_loss}
\end{figure*}

\begin{figure*}
\centering
\subfigure[$P_{c-avail}$ versus  $\lambda_f$]
{\includegraphics[trim= 2cm 0.05cm 3cm 0.4cm, height = 5.3cm,width = 0.5\textwidth]{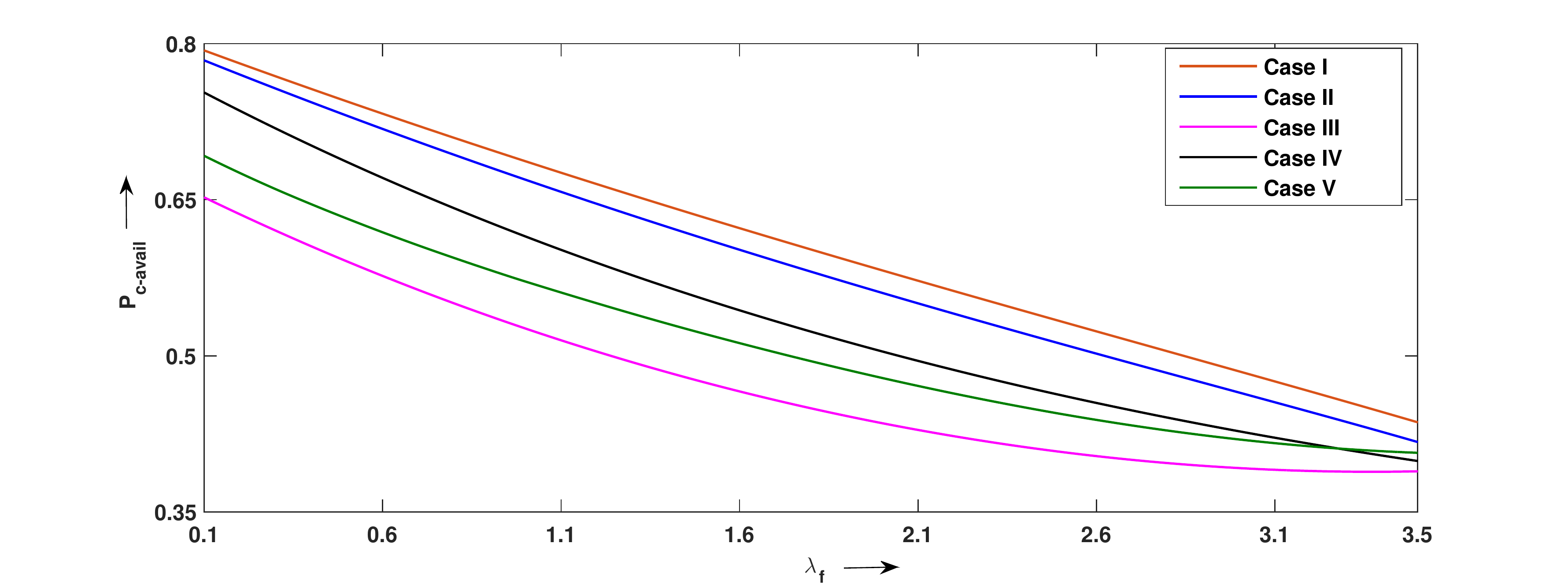}
\label{fig:reliability_failure}}%
\subfigure[$P_{c-avail}$ versus  $\mu_r$]
{\includegraphics[trim= 2.5cm 0.2cm 2cm 0.5cm, clip=true, height = 5.3cm,width = 0.5\textwidth]{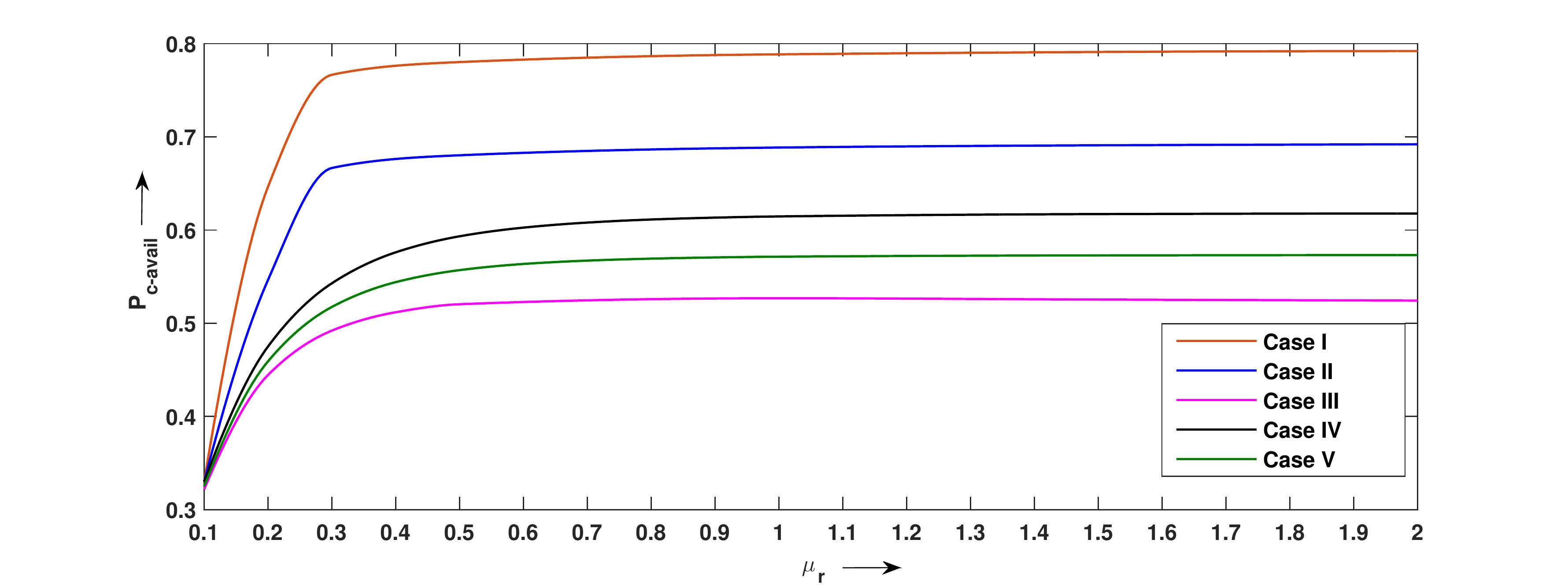}
\label{fig:reliability_repair}}%
\caption{Dependence of the probability of channel availability in busy state $P_{c-avail}$ over failure intensity $\lambda_f$ and repair intensity $\mu_r$. }
\label{fig:reliability}
\end{figure*}

\noindent \textbf{Illustration 3:} The main propose of this illustration is to analyse how the  the probability of  call loss due to the occurrence of channel failure ($P_{loss}^{c-failure}$) and the probability of channel availability in busy state ($P_{c-avail}$)
are affected by the channel failure and repair intensity.
\begin{itemize}
    \item In Figures. \ref{fig:P_loss} \subref{fig:loss_failure} and \ref{fig:P_loss} \subref{fig:loss_repair},  the impacts of failure intensity $\lambda_f$ and repair intensity $\mu_r$ over $P_{loss}^{c-failure}$ are shown, respectively. It is clear from Figure. \ref{fig:P_loss} \subref{fig:loss_failure} that an increment in the value of $\lambda_f$ causes a rapid increment in the value of $P_{loss}^{c-failure}.$  This particular finding could be elucidated by noticing that, for the fixed number of total channels, the number of failed channels will keep on increasing as $\lambda_f$ increases. Thus, the probability of losing  call will increase.  The vice versa effect can be observed in the Figure. \ref{fig:P_loss} \subref{fig:loss_repair} where $P_{loss}^{c-failure}$ decreases as $\mu_r$ increases, which is expected. Observe that in both scenarios, $P_{loss}^{c-failure}$ has the highest value for Case I among all other cases. This verifies the fact that the disregard of correlation and channel failures might create huge flaws in the estimation of performance of the system.
    
    \item  Figures. \ref{fig:reliability} \subref{fig:reliability_failure} and \ref{fig:reliability} \subref{fig:reliability_repair} depict the dependency of  $P_{c-avail}$  over $\lambda_f$ and $\mu_r$, respectively.
    As expected, $P_{c-avail}$ decreases with the increasing value of $\lambda_f$  whereas increases with the increasing value of  $\mu_r$. If the channel failure occur rapidly in the system, it will decrease the availability of working channels in the system, which is  obvious in real-life scenario. Another interesting observation is that after a certain  value of $\lambda_f$ and $\mu_r$, the value of $P_{c-avail}$ becomes almost invariable which explains the steady-state of the understudy model.  For all the cases involving different coefficient of correlation and coefficient of variations, we observe that consideration of higher correlation and variation yields a higher value for $P_{c-avail}$.
    \end{itemize}
 
\begin{figure*}
\centering
\subfigure[$\theta_r^{succ}$ versus  $\theta$]
{\includegraphics[trim= 2cm 0.1cm 3cm 0.4cm, height = 5.3cm,width = 0.5\textwidth]{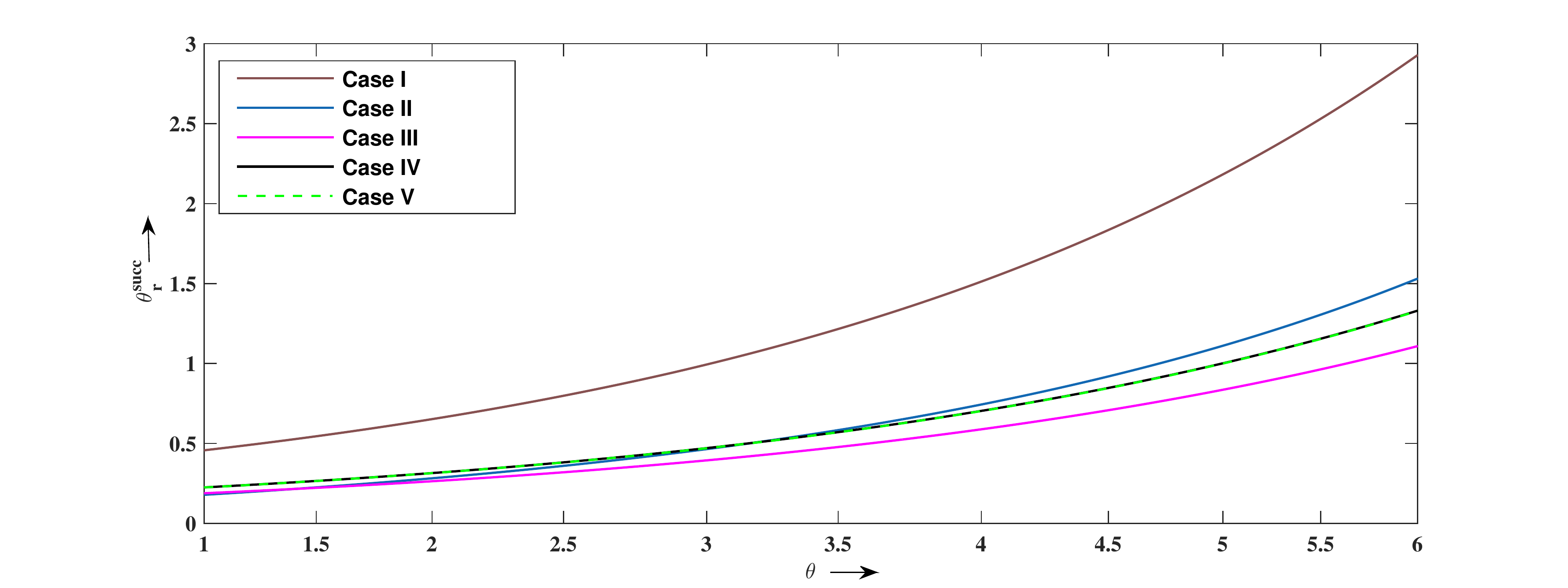}
\label{fig:rate_capture_channel}}%
\subfigure[$P_{leave}^{no-service}$ versus $\theta$]
{\includegraphics[trim= 2.5cm 0.05cm 2cm 0.5cm, clip=true, height = 5.3cm,width = 0.5\textwidth]{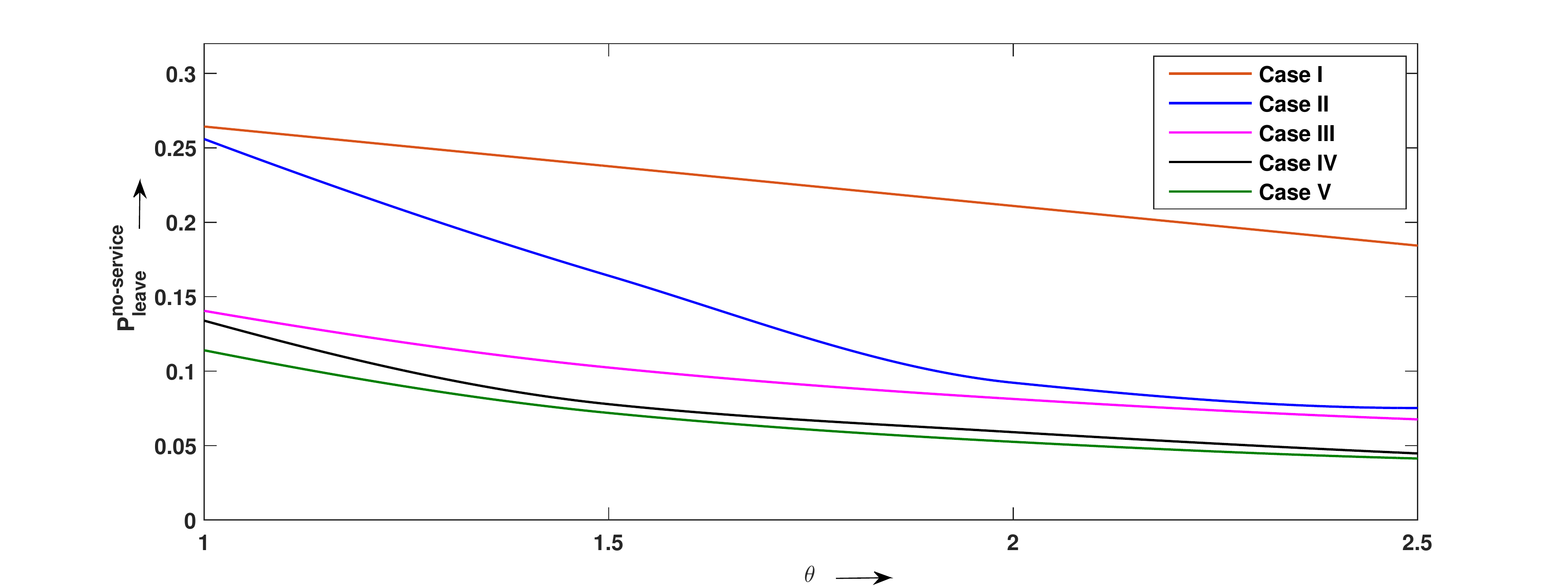}
\label{fig:impatience_retrial}}%
\caption{Dependence of the intensity by which a retrial call is successfully connected to an available channel $\theta_r^{succ}$ and    the probability of a retrial call completing the cell residence time  without getting the service $P_{leave}^{no-service}$ over retrial intensity $\theta$. }
\label{fig:retrial_rate}
\end{figure*}

\noindent \textbf{Illustration 4:} Here, we will focus over the effect of retrial intensity ($\theta$) over two performance measures, the intensity by which a retrial call is successfully connected to an available channel  ($\theta_r^{succ}$) and the probability that the retrial call  will complete the cell residence time without getting the service ($P_{leave}^{no-service}$). 
\begin{itemize}
   \item  Figure. \ref{fig:retrial_rate} \subref{fig:rate_capture_channel} demonstrates that $\theta_r^{succ}$ behaves as an increasing function of the retrial intensity, which is obvious also. This behaviour can easily be described as an increase in  retrial intensity increases the probability of getting connection for a retrial call, which results in the increased value of $\theta_r^{succ}$. Moreover, there can be observed a significant difference in the values of $\theta_r^{succ}$ for Case I in contrast to other cases which describes how the consideration of correlation and variation in the system can impact system performance.
 \item  Figure. \ref{fig:retrial_rate} \subref{fig:impatience_retrial} depicts  that the measure, $P_{leave}^{no-service}$  decreases as  retrial intensity increases. If the retrial call retries with a faster rate, it will increase the probability of getting the connection, so the probability of a retrial call leaving the system without obtaining the service will decrease. Similarly, Case I yields a larger value for $P_{leave}^{no-service}$ in comparison to other cases. This result again confirms the importance of correlation and variation in the cellular networks. 
\end{itemize}
In the next subsection, a cost optimization problem concerning expected cost per unit time for the proposed model will be demonstrated on the basis of the system performance measures mentioned in Section \ref{section4}.
 

  \begin{figure*}
\centering
\subfigure[$f$ versus  $\mu$ and $\mu_r$]
{\includegraphics[trim= 0.1cm 0.05cm 0.01cm 0.001cm, height = 5.5cm,width = 0.55\textwidth]{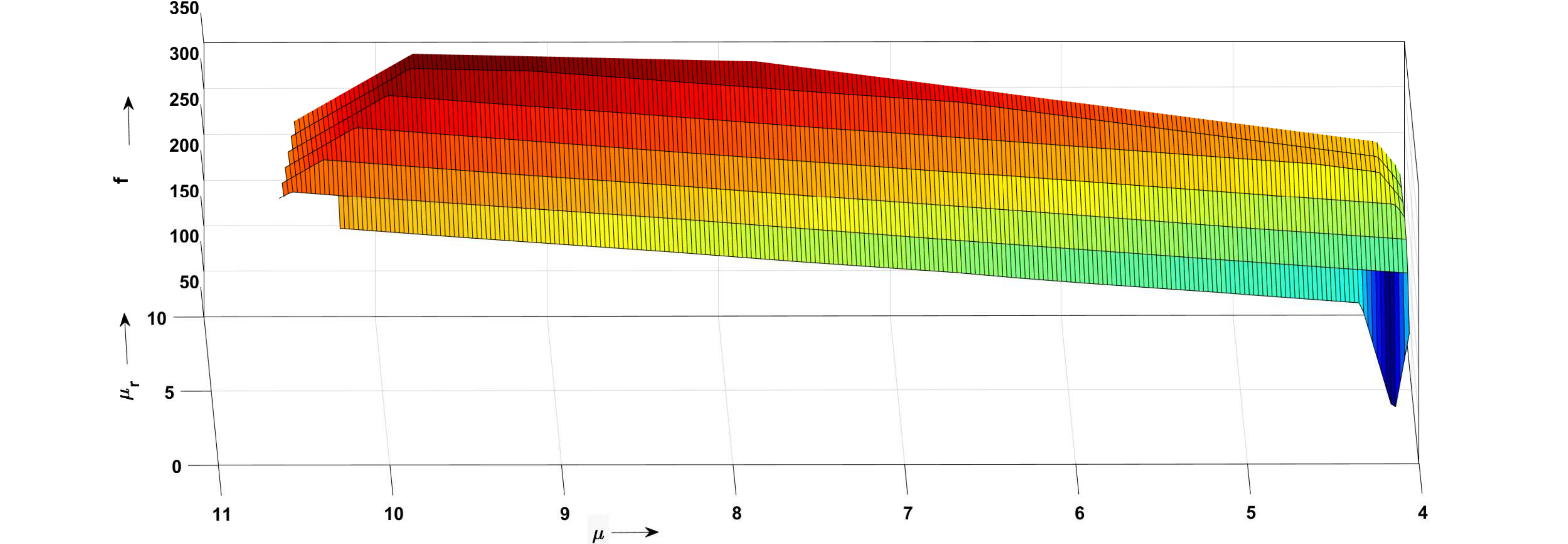}
\label{fig:cost_fcn}}%
\subfigure[$f$ versus  iteration]
{\includegraphics[trim= 0.3cm 0.1cm 0.1cm 0.1cm,  height = 5.8cm,width = 0.5\textwidth]{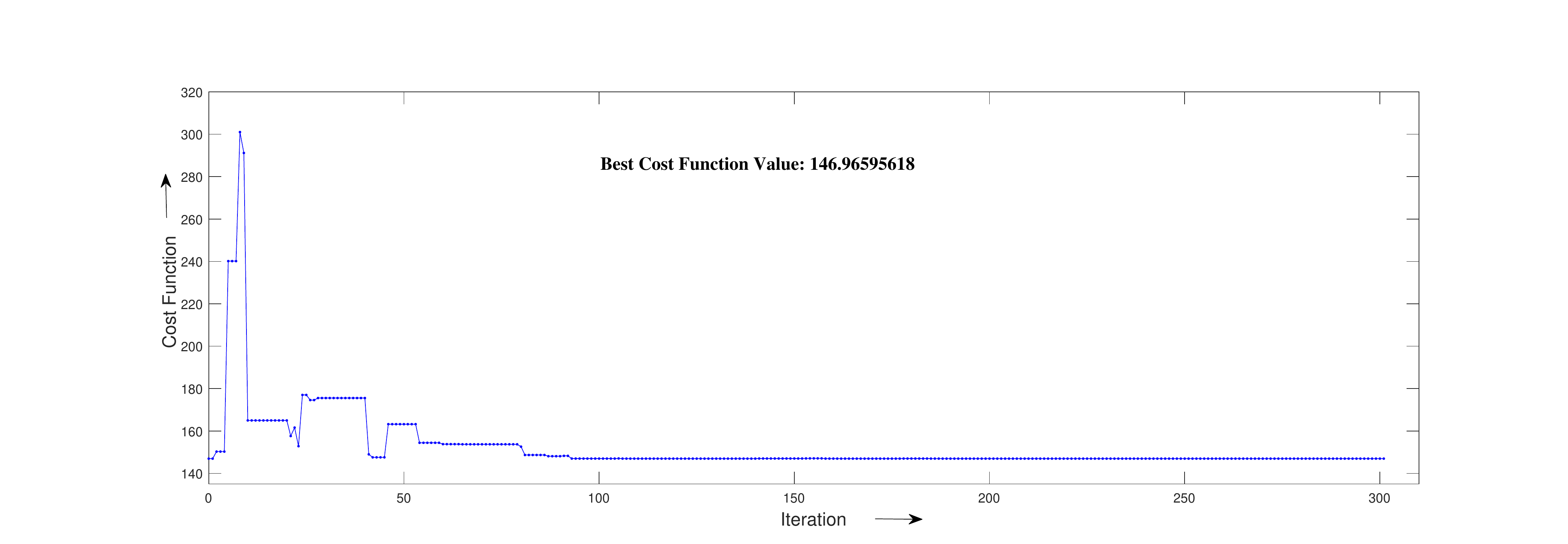}
\label{fig:cost_fcn_iteration}}%
\caption{Dependence of the cost function $f$ over number of iterations, service intensity $\mu$ and retrial intensity $\mu_r$ for $\lambda =2$, $\lambda_f =10.$ }
\label{fig:Cost_fcn}
\end{figure*}
\subsection{Cost Optimization Problem}\label{section6}

For the service provider, service cost and repair cost (in case of  channel failures) are essential factors to determine the net profit or loss. Given the large number of channels in use today, the expenditure on the service and repair process should be considered separately, as the failed channels require more cost to get repair. There can be many other factors also which are responsible for the expenditure of a cellular network such as labor cost, R \& D costs, depreciation cost, expenditure on retrial customers, etc. In this study, we propose an approximated cost function which will minimize the expenditure  for the service provider  by obtaining the optimal value of the service intensity $\mu$ and repair intensity $\mu_r$.  In order to describe the cost function, following cost factors are defined\\
$C_{EB}$ $:=$ cost per unit time for  one channel being in service,\\
$C_{EN}$ $:=$ cost per unit time for one channel being failed,\\
$C_S$ $:=$ cost per unit time of providing service intensity $\mu$,\\
$C_R$ $:=$ cost per unit time of providing repair intensity $\mu_r$.\\

On the basis of cost factors mentioned above, the unconstrained cost optimization problem is defined as follows

$\begin{array}{ll}
&\textrm{Min }  ~f(\mu, \mu_r)  = C_{EB}E\!B + C_{EN}E\!N + C_S \mu + C_R \mu_r,\\
&~~~~~~~~~\mu,\mu_r \geq 0.
\end{array}$\\
Here, $E\!B$ and $E\!N$ are number of busy channels and number of failed channels, respectively.  
Due to the highly complex and non-linear structure of cost function, a heuristic algorithm named, Simulated Annealing ($S\!A
$) method  has been applied to obtain an approximate solution. Though, the convergence rate of $S\!A$ method is comparatively slower than other heuristic methods, yet it successfully obtain the global optimum solution without having a prior knowledge about the differentiablility of the objective function. The algorithm for $S\!A$ method is presented as follows\\

\textbf{Algorithm :}
\begin{itemize}
	\item Step 1: Fix the parameters $\mathpzc{S}$, $\mathpzc{G}$, $\lambda_{\mathcal{N}}$, $\lambda_{\mathcal{H}}$, $\lambda_{f}$ and $\theta$  for the computation of the cost function.
	\item Step 2: Initialize $s$ as the initial state. Generate $s^{'}$ $= s + \Delta s$ a neighbor state of $s.$
	\item Step 3: Compute $\Delta f = f(s) - f(s^{'}).$
	\item Step 4: Generate $P = e^{(-\Delta f/T)}$, the acceptance probability of $s.$ Where $T$ is  the temperature parameter which is  evaluated randomly by computing the mean of cost functions for differenet values.
	\item Step 5: Generate $R$ = $U(0,1)$, the acceptance probability of $s^{'}$.
	\item Step 6: If $\Delta f(s) < 0$; $s$ is the actual state else if $R>P$; $s^{'}$ is the actual state.
	\item Step 7: Repeat steps 2-6 until $\Delta f < 10^{-5}$.
	\item Step 8: Output optimum solution $\mu^{*}$, $\mu_r^{*}$ and $f$.
\end{itemize}

For the computation propose, the values of cost factors are set as
 $C_{EB}=$ 10/channel; $C_{EN}=$ 15/channel; $C_S =$ 15/unit time; $C_R =$ 20/unit time. The value of other parameters are same as proposed in Section \ref{section5}. All results were obtained by MATLAB software, which were run on a computer with Intel Core i7-6700 3.40GHz CPU and 8 GB of RAM.
Table \ref{tab:my_label1} exhibits the optimal values of $\mu^*$, $\mu_r^*$ and $f$  obtained by applying $S\!A$ method for different combination of parameters $\lambda$ and $\lambda_f$. Figures.  \ref{fig:Cost_fcn} \subref{fig:cost_fcn} and \ref{fig:Cost_fcn} \subref{fig:cost_fcn_iteration} depict the behaviour of cost function $f$ with respect to $\mu$ and $\mu_r$. The value of cost function increases  as the arrival intensity and failure intensity increase.  Figure \ref{fig:Cost_fcn} \subref{fig:cost_fcn_iteration} shows that the convergence of $S\!A$ method is slow but it provides the global optimum value for the cost function. 
\begin{table}
	\centering
	\scalebox{0.8}{
	\begin{tabular}{llllllll}
		
		\hline
		$  \lambda_f = 10$ & & & & & & &\\
		\hline
		$\lambda$ & 0.5 & 0.75 & 1 & 1.25 & 1.5 & 1.75 & 2 \\
		$\mu^*$ & 4.00001015 &4.000064 &4.0000144 &4.0000528 & 4.0000534& 4.00005119 &4.00000425
 \\
		$\mu_r^*$ & 2.50000581 & 2.50000173 & 2.5000018 & 2.50001554 & 2.5 & 2.50000782 & 2.5000001
 \\
		$f$ & 122.0414 & 128.7040 &   132.7520 & 137.23516 & 141.102268 & 144.34092& 146.9659562

		\\
		\hline
		$  \lambda_f = 9$ & & & & & & &\\
		\hline
		$\lambda$ & 0.5 & 0.75 & 1 & 1.25 & 1.5 & 1.75 & 2 \\
		$\mu^*$ & 4.0000173 &4.0000491 &4.0000519 &4.0001013& 4.0000389& 4.0000023& 4.00000438\\
		$\mu_r^*$ & 2.50000751 & 2.5000053 & 2.5000013 & 2.50000342 & 2.50000296 & 2.5000047 & 2.5\\
		$f$ & 121.6584 & 127.23255 & 132.32399 & 136.88873 & 140.996749 & 144.3108486 & 147.1484662
		\\
		\hline
	$  \lambda_f = 8$ & & & & & & &\\
		\hline
		$\lambda$ & 0.5 & 0.75 & 1 & 1.25 & 1.5 & 1.75 & 2 \\
		$\mu^*$ & 4.00002823 &4.00005668 &4.00001281 & 4.00000002& 4.00000357& 4& 4.0000227\\
		$\mu_r^*$ & 2.50000726 & 2.50001473 & 2.50000117 & 2.50000006 & 2.50000365 & 2.50000661 & 2.5\\
		$f$ & 121.32840 & 126.832031 & 131.9184364 & 136.533562 & 140.63595768 & 144.19533571 & 147.2047622
		\\
		\hline
		$  \lambda_f = 7$ & & & & & & &\\
		\hline
		$\lambda$ & 0.5 & 0.75 & 1 & 1.25 & 1.5 & 1.75 & 2 \\
		$\mu^*$ &  4.00000131&4.00000039 & 4.00000327 &4.000000537 & 4.0000181 & 4.00002120& 4.00002239\\
		$\mu_r^*$ & 2.50000072 & 2.500000188 & 2.50000004 & 2.5 & 2.50000083 & 2.5 & 2.50000776\\
		$f$ & 121.04098838 & 126.471810 & 131.95865 & 136.18451443 & 140.35815138 & 144.025 & 147.573589
		\\
		\hline
		$  \lambda_f = 6$ & & & & & & &\\
		\hline
		$\lambda$ & 0.5 & 0.75 & 1 & 1.25 & 1.5 & 1.75 & 2 \\
		$\mu^*$ & 4.00000043 & 4.0000154 & 4.00000506 & 4.0000106 & 4.00000106 & 4.00000193 & 4.00000009\\
		$\mu_r^*$ & 2.50000309 & 2.5 & 2.50000463 & 2.50000782 & 2.50000782 & 2.50000578 & 2.50000654\\
		$f$ & 120.79014934 & 126.14890546 & 131.18994940 & 135.8479802 & 140.07201506 & 143.822826 & 147.0816232
		\\
		\hline
	\end{tabular}}
	\caption{Optimal values of $\mu^*$, $\mu_r^*$ and $f$ for different values of $\lambda$ and $\lambda_f$ by applying $S\!A$ method.}
	\label{tab:my_label1}
\end{table}

\section{Conclusions}\label{section7}

With the tremendous growth of customers in the cellular networks, the reliability of the system has become an  essential factor to determine the  performance of the system. Therefore, this study presents an integrated performability model with {\it{MAP}} input flow and  $\textrm{\it PH}$ distributed service times with different service rates.  The incoming  calls are prioritized using guard channel policy. Due to the brief span of inter-retrial times in comparison to service times, a more generalized approach, $\textrm{\it PH}$ distributed retrial times  is used so that the performance of the system is not over or under estimated.  The computational complexities, arising due to the consideration of $\textrm{\it PH}$ distributed retrial times have been tackled by applying the truncation method. The  steady-state distribution of the $\textrm{\it LDQBD}$ process is analyzed using the $\textrm{\it MAM}$. Through the numerical illustration of various performance measures, it has been shown that disregard of correlation and variation for arrival, service, and retrial processes can create significant flaws while approximating the performance of the system. Since the proposed system is subjected to channel failures, this leads to the formulation of cost optimization problem for the service provider. It is essential for the operator to control service and repair intensities in a proper  way to optimize the operating costs. This cost minimization problem has been solved by implementing $\textrm{\it SA}$ method that provides the optimal values of service and repair intensities. This proposed model can be extended to include  various versatile arrival and service processes to reflect a more realistic scenario i.e., batch Markovain arrival process, Markovian service process.

\end{document}